\documentclass[a4paper,12pt]{article}
\usepackage{setspace,graphics}
\usepackage[dvips]{epsfig}

\usepackage{amssymb,epsfig,euscript,array,cite}

\newcounter{multieqs}

\newcommand{\be}{\begin{equation}}
\newcommand{\ee}{\end{equation}}
\newcommand{\eq}[1]{(\ref{#1})}

\newcommand{\bm}[1]{\mbox{\boldmath $#1$}}

\def\bd{\begin{document}}
\def\ed{\end{document}}
\def\nn{\nonumber}
\def\bea{\begin{eqnarray}}
\def\eea{\end{eqnarray}}
\def\obar{\overline}
\let\bm=\bibitem

\def\npb#1#2#3{Nucl. Phys. {\bf{B#1}} #3 (#2)}
\def\plb#1#2#3{Phys. Lett. {\bf{#1B}} #3 (#2)}
\def\prl#1#2#3{Phys. Rev. Lett. {\bf{#1}} #3 (#2)}
\def\prd#1#2#3{Phys. Rev. {D \bf{#1}} #3 (#2)}
\def\cmp#1#2#3{Comm. Math. Phys. {\bf{#1}} #3 (#2)}
\def\cqg#1#2#3{Class. Quantum Grav. {\bf{#1}} #3 (#2)}
\def\nppsa#1#2#3{Nucl. Phys. B (Proc. Suppl.) {\bf{#1A}}#3 (#2)}
\def\ap#1#2#3{Ann. of Phys. {\bf{#1}} #3 (#2)}
\def\ijmp#1#2#3{Int. J. Mod. Phys. {\bf{A#1}} #3 (#2)}
\def\rmp#1#2#3{Rev. Mod. Phys. {\bf{#1}} #3 (#2)}
\def\mpla#1#2#3{Mod. Phys. Lett. {\bf A#1} #3 (#2)}
\def\jhep#1#2#3{J. High Energy Phys. {\bf #1} #3 (#2)}
\def\atmp#1#2#3{Adv. Theor. Math. Phys. {\bf #1} #3 (#2)}

\newcommand{\EQ}[1]{\begin{equation} #1 \end{equation}}
\newcommand{\AL}[1]{\begin{subequations}\begin{align} #1 
\end{align}\end{subequations}}
\newcommand{\SP}[1]{\begin{equation}\begin{split} #1 \end{split}\end{equation}}
\newcommand{\ALAT}[2]{\begin{subequations}\begin{alignat}{#1} #2 
\end{alignat}\end{subequations}}
\def\beqa{\begin{eqnarray}} 
\def\eeqa{\end{eqnarray}} 
\def\beq{\begin{equation}} 
\def\eeq{\end{equation}} 

\def\N{{\cal N}}
\def\sst{\scriptscriptstyle}
\def\thetabar{\bar\theta}
\def\Tr{{\rm Tr}}

\def\a{\alpha}          \def\da{{\dot\alpha}}
\def\b{\beta}           \def\db{{\dot\beta}}
\def\c{\gamma}  \def\C{\Gamma}  \def\cdt{\dot\gamma}
\def\d{\delta}  \def\D{\Delta}  \def\ddt{\dot\delta}
\def\e{\epsilon}                \def\vare{\varepsilon}
\def\f{\phi}    \def\F{\Phi}    \def\vvf{\f}
\def\h{\eta}
\def\k{\kappa}
\def\l{\lambda} \def\L{\Lambda} \def\la{\lambda}
\def\m{\mu}     \def\n{\nu}
\def\o{\omega}
\def\p{\pi}     \def\P{\Pi}
\def\r{\rho}
\def\s{\sigma}  \def\S{\Sigma}
\def\t{\tau}
\def\th{\theta} \def\Th{\Theta} \def\vth{\vartheta}
\def\X{\Xeta}
\def\z{\zeta}

\def\cA{{\cal A}} \def\cB{{\cal B}} \def\cC{{\cal C}}
\def\cD{{\cal D}} \def\cE{{\cal E}} \def\cF{{\cal F}}
\def\cG{{\cal G}} \def\cH{{\cal H}} \def\cI{{\cal I}}
\def\cJ{{\cal J}} \def\cK{{\cal K}} \def\cL{{\cal L}}
\def\cM{{\cal M}} \def\cN{{\cal N}} \def\cO{{\cal O}}
\def\cP{{\cal P}} \def\cQ{{\cal Q}} \def\cR{{\cal R}}
\def\cS{{\cal S}} \def\cT{{\cal T}} \def\cU{{\cal U}}
\def\cV{{\cal V}} \def\cW{{\cal W}} \def\cX{{\cal X}}
\def\cY{{\cal Y}} \def\cZ{{\cal Z}}

\def\ua{\underline{\alpha}}
\def\ub{\underline{\phantom{\alpha}}\!\!\!\beta}
\def\uc{\underline{\phantom{\alpha}}\!\!\!\gamma}
\def\um{\underline{\mu}}
\def\ud{\underline\delta}
\def\ue{\underline\epsilon}
\def\una{\underline a}\def\unA{\underline A}
\def\unb{\underline b}\def\unB{\underline B}
\def\unc{\underline c}\def\unC{\underline C}
\def\und{\underline d}\def\unD{\underline D}
\def\une{\underline e}\def\unE{\underline E}
\def\unf{\underline{\phantom{e}}\!\!\!\! f}\def\unF{\underline F}
\def\unm{\underline m}\def\unM{\underline M}
\def\unn{\underline n}\def\unN{\underline N}
\def\unp{\underline{\phantom{a}}\!\!\! p}\def\unP{\underline P}
\def\unq{\underline{\phantom{a}}\!\!\! q}
\def\unQ{\underline{\phantom{A}}\!\!\!\! Q}
\def\unH{\underline{H}}

\def\As {{A \hspace{-6.4pt} \slash}\;}
\def\bs {{b \hspace{-6.4pt} \slash}\;}
\def\Ds {{D \hspace{-6.4pt} \slash}\;}
\def\ds {{\del \hspace{-6.4pt} \slash}\;}
\def\ss {{\s \hspace{-6.4pt} \slash}\;}
\def\ks {{ k \hspace{-6.4pt} \slash}\;}
\def\ps {{p \hspace{-6.4pt} \slash}\;}
\def\pas {{{p_1} \hspace{-6.4pt} \slash}\;}
\def\pbs {{{p_2} \hspace{-6.4pt} \slash}\;}

\def\Fh{\hat{F}}
\def\Vh{\hat{V}}
\def\Xh{\hat{X}}
\def\ah{\hat{a}}
\def\xh{\hat{x}}
\def\yh{\hat{y}}
\def\ph{\hat{p}}
\def\xih{\hat{\xi}}

\def\psit{\tilde{\psi}}
\def\Psit{\tilde{\Psi}}
\def\tht{\tilde{\th}}
 
\def\At{\tilde{A}}
\def\Qt{\tilde{Q}}
\def\Rt{\tilde{R}}

\def\st{\tilde{s}}
\def\ft{\tilde{f}}
\def\pt{\tilde{p}}
\def\qt{\tilde{q}}
\def\vt{\tilde{v}}

\def\delb{\bar{\partial}}
\def\bz{\bar{z}}
\def\bD{\bar{D}}
\def\bB{\bar{B}}

\def\bk{{\bf k}}
\def\bl{{\bf l}}
\def\bp{{\bf p}}
\def\bq{{\bf q}}
\def\br{{\bf r}}
\def\bx{{\bf x}}
\def\by{{\bf y}}
\def\bR{{\bf R}}
\def\bV{{\bf V}}

\def\R{{\mathbb R}}
\def\C{{\mathbb C}}
\def\N{{\mathbb N}}
\def\Z{{\mathbb Z}}
\def\one{\mbox{1 \kern-.59em {\rm l}}}

\def\bit{\begin{itemize}}
\def\eit{\end{itemize}}

\def\({\left(}
\def\){\right)}
\def\diag{\mbox{diag}}
\def\refeq#1{(\ref{#1})}

\def\tens{\otimes}
\def\Pint{\int\kern-11.5pt-\kern7pt}

\def\d{\delta}\def\D{\Delta}\def\ddt{\dot\delta}

\def\pa{\partial} \def\del{\partial}
\def\xx{\times}
\def\uno{\mbox{1 \kern-.59em {\rm l}}}  

\def\trp{^{\top}}
\def\inv{^{-1}}
\def\dag{{^{\dagger}}}
\def\pr{^{\prime}}

\def\rar{\rightarrow}
\def\lar{\leftarrow}
\def\lrar{\leftrightarrow}

\def\one{1\!\!1\,\,}
\def\im{\imath}
\def\jm{\jmath}

\newcommand{\tr}{\mbox{tr}}
\newcommand{\slsh}[1]{/ \!\!\!\! #1}

\def\vac{|0\rangle}
\def\lvac{\langle 0|}

\def\hlf{\frac{1}{2}}
\def\ove#1{\frac{1}{#1}}

\def\Box{\square}
\def\ZZ{\mathbb{Z}}
\def\CC#1{({\bf #1})}
\def\bcomment#1{}
\def\bfhat#1{{\bf \hat{#1}}}
\def\VEV#1{\left\langle #1\right\rangle}

\newcommand{\ex}[1]{{\rm e}^{#1}} \def\ii{{\rm i}}

\begin{document}

\begin{flushright}
LMU-TPW 03-05 \\
\end{flushright}

\vspace{20pt}

\begin{center}

{\Large \bf Quantized Gauge Theory on the Fuzzy Sphere \\ 
as Random Matrix Model} 
\vspace{30pt}
 
{\bf Harold Steinacker}
 
\vspace{15pt}

{\small \em Institut f\"ur theoretische Physik \\
Ludwig--Maximilians--Universit\"at M\"unchen \\
Theresienstr.\ 37, D-80333 M\"unchen, Germany}

{\small {\it Email:} 
\tt harold.steinacker@physik.uni-muenchen.de}

\vspace{30pt}
{\bf Abstract}
\end{center}

$U(n)$ Yang-Mills theory on the fuzzy sphere $S^2_N$ is quantized using 
random matrix methods. The gauge theory is formulated  
as a matrix model for a single Hermitian 
matrix subject to a constraint, and a potential with two 
degenerate minima. This allows to reduce the path integral 
over the gauge fields to an integral over eigenvalues,
which can be evaluated for large $N$. 
The partition function of $U(n)$ Yang-Mills theory on the classical
sphere is recovered in the large $N$ limit,
as a sum over instanton contributions.
The monopole solutions are found explicitly.

\vspace{0.5cm}

\setcounter{page}0
\thispagestyle{empty}
\newpage

\begin{spacing}{.3}
{
\noindent\rule\textwidth{.1pt}            
   \tableofcontents 
\vspace{.6cm}
\noindent\rule\textwidth{.1pt}
}
\end{spacing}

\section{Introduction}

Gauge theories provide the best known description of the fundamental
forces in nature. At very short distances however, physics is not known, 
and it is plausible that spacetime is 
quantized below some scale. This idea has been contemplated 
for quite some time, and received a boost recently
due to the discovery that string theory
naturally leads to noncommutative gauge theories under suitable 
conditions, as explained in \cite{SW}.
Gauge theory on noncommutative spaces has been the 
subject of much research activity in recent years, see e.g.
\cite{douglas} for a review.

There is one major problem with most models of noncommutative 
gauge theories: their quantization is very difficult.
A direct quantization leads to difficulties related to the notorious 
UV/IR mixing \cite{seiberg}, see also \cite{wulke-new} for
some recent developments.
On the other hand, the use of the Seiberg-Witten map \cite{SW,wess}, 
which allows a formulation in terms of commutative 
quantities, yields Lagrangians which become increasingly complicated 
at each order in the deformation parameter. 
This seems to rule out perturbative quantization \cite{wulki-non}.

The motivation behind this paper is to try to develop 
{\em new tools} for the quantization of gauge theories, 
taking advantage of noncommutativity.
The idea is to make use of one very fascinating feature
of gauge theory on (some) noncommutative spaces:
It is possible to formulate the gauge theories 
in terms of Lagrangians which have {\em no} derivatives. 
Rather, the dynamical variables are essentially matrices $B_i$,
and the action is the trace of products of these matrices. 
Hence these gauge theories
are {\em matrix models}. The gauge transformations have the form
$B_i \to U^{-1} B_i U$ for unitary matrices $U$.
The kinetic term is generated upon a 
shift $B_i = X_i + A_i$, and the $A_i$ become the usual gauge fields
in the commutative limit. 
This is very interesting for the quantization, because the 
path integral can now 
be defined simply as the integral over the matrices $B_i$,
as in a random matrix model. 
A promising strategy is then to first do the quantization in terms
of the $B_i$ fields, and then go to the ``classical'' variables $A_i$.
This is a bit in the spirit of \cite{cole-wein}.

In general of course, things are still complicated: The actions are 
multi-matrix models with nontrivial interactions, and the 
integration over the $B_i$
is highly nontrivial. It must be so, since they describe
a nontrivial quantum field theory. Moreover, 
the matrices are infinite-dimensional for most spaces
(such as for $\R^n_\theta$). This latter problem does not
occur on the so-called fuzzy spaces, in particular the fuzzy sphere
$S^2_N$ \cite{madore}. This quantum space is characterized by a 
deformation parameter $\frac 1N$ which measures the size of 
``Planck cells'', and reduces to the classical sphere
for $N \to \infty$. Moreover the rotation invariance under 
$SU(2)$ is maintained, hence $S^2_N$ seems particularly well suited to 
explore this idea.

In this paper, we will show that for pure $U(n)$ Yang-Mills theory
on the fuzzy sphere, the quantization can be carried
out completely by integrating over the matrices $B_i$. This is achieved
by collecting the $B_i$ into a single hermitian
matrix, subject to a constraint.
Of course, Yang-Mills theory on $S^2$ is a rather simple field
theory with no propagating degrees of freedom; however it does have
nontrivial monopole sectors, and its quantization is not entirely trivial.
We will calculate the partition function for $U(n)$ Yang-Mills theory on 
$S^2_N$ in the large $N$ limit, and 
recover the known result \cite{migdal,rusakov} 
for the partition function on the classical
sphere. Corrections of order $\frac 1N$ could be calculated in principle, 
but we do not attempt this here. The main message is 
the applicability of completely new methods to noncommutative gauge
theory, and hence to their commutative limit. Moreover, our result 
strongly suggests that the
``commutative limit'' of pure gauge theory on the fuzzy sphere is smooth,
which is not obvious in view of the UV/IR mixing effects
in noncommutative field theories \cite{seiberg}.
This was also found recently on the quantum torus \cite{szabo},
however with very different methods.

Another important message is that in the approach developed here, the
construction of gauge theory on (some)
noncommutative spaces can be {\em simpler} than on a classical space.
In particular, there is no need to introduce nontrivial fiber bundles,
connections and other mathematical structures in our approach: 
the monopole sectors arise automatically in a very simple way, 
and reproduce the correct classical limit. 
We explicitly calculate the gauge fields for all monopole configurations.

This paper is organized as follows. After a brief review of the fuzzy sphere
and the most basic facts about matrix models, we present in 
Section \ref{subsec:new} the particular potential to be used in this paper.
We then show that its minima define a fuzzy sphere, and 
the fluctuations become gauge fields on this fuzzy sphere after imposing
a suitable constraint. In Section \ref{subsec:monopole} the monopole
sectors of the $U(1)$ case are identified, and the gauge field for the
monopoles is calculated explicitly. We then generalize the construction 
to the $U(n)$ case in Section \ref{sec:nonabelian}, which amounts simply 
to taking larger matrices. Section \ref{sec:pathintegral} 
contains the calculation of the path integral, which is the main 
application of our construction. Finally we make some 
simple observations on
symmetries, correlation functions, and show how a small modification 
leads to gauge theory on the $q$-deformed fuzzy sphere.
The technical part of the path integral calculation is postponed to the 
appendix. 
In general, the focus is on explicit calculations, 
keeping the formal mathematics to a minimum.
The hope is that noncommutative field theory in general and 
at least some of the techniques developed here
will eventually be useful for physics.

\section{The basic fuzzy sphere}
\label{sec:basic}

We start by recalling the definition of fuzzy sphere \cite{madore,grosse1}
in order to fix
our conventions. The algebra of
functions on the fuzzy sphere is the finite algebra $S_N^2$
generated by Hermitian operators
$x_i= (x_1, x_2, x_3)$ satisfying the defining relations
\bea
[x_i, x_j] = i \L_N \e_{ijk} x_k, \label{def1}\\
x_1^2 + x_2^2 +x_3^2  = R^2. \label{def2}
\eea
The noncommutativity parameter $\L_N$ is of dimension length, and
can be taken positive. The radius $R$ is quantized in units of $\L_N$ by
\be\label{def3}
\frac {R}{\L_N} = \sqrt{\frac{N^2-1}{4}} \; ,
\quad \mbox{$N = 1,2,\cdots$ }
\ee
This quantization can be easily understood. Indeed \eq{def1} is
simply the Lie algebra $su(2)$, whose irreducible representation
have dimension $N$. The Casimir
of the $N$-dimensional representation is quantized, and related to $R^2$ by
\eq{def2} and \eq{def3}.
Thus the fuzzy sphere is characterized by its radius $R$ and the
``noncommutativity parameters'' $N$ or $\L_N$.
The algebra of ``functions'' $S_N^2$ is simply the
algebra $Mat(N)$ of $N \times N$ matrices.
It is covariant under the adjoint action of $SU(2)$, under which it
decomposes into the irreducible representations with dimensions
$(1) \oplus (3) \oplus (5) \oplus ... \oplus (2N-1)$.
The integral of a function $f \in S_N^2$ over the fuzzy sphere is given by
\be
R^2 \int f(x)  = \frac{4 \pi R^2}{N} \Tr[ f(x)],
\ee
where we have introduced $\int$, the integral over the fuzzy sphere with
unit radius. 
It agrees with the integral $\int d \Omega$  on $S^2$ in the large $N$ limit.
Invariance of the integral under  the rotations
$SU(2)$ amounts to invariance of the trace under adjoint action.
It is convenient to introduce the dimensionless coordinates
\beq \label{lx}
\l_i = x_i/\L_N
\eeq
which satisfy
\beq \label{FS-l}
\vare^{ij}_k \l_i \l_j = i \l_k,
 \quad \l_i \l^i  = \frac{N^2-1}{4}.
\eeq
The $\l_i$ form a $N$-dimensional representation of $SU(2)$, which is given
explicitly in Appendix A for convenience.
Noting that 
$[\l_i,x_j] = i \vare_{ijk} x^k$,
it follows that the rotation operators $J_i$ act 
on functions $f \in S_N^2$ as
\beq
J_i f = [\l_i, f].
\label{J-def}
\eeq
One can now write down actions for scalar fields, such as 
\beq
S_0 = \int \frac 12\; \Phi (\Delta + \m^2) \Phi + 
    \frac{g}{4 !} \Phi^4
\eeq
were $\Phi$ is a Hermitian matrix,
and $\Delta = \sum J_i^2$ is the Laplace operator.
For gauge fields, the ``correct'' action is less obvious because the
gauge fields have a priori 3 components (because there are 3 independent
one-forms on $S^2_N$, \cite{madore}), and it is not obvious how to get rid of 
the normally unwanted 3rd component.
Several slightly different approaches have been pursued in the literature
\cite{madore,grossegauge,balagauge,klimcik,watamura}. 
Alternatively, if one keeps the 3rd component
which is essentially a scalar field, one finds actions which 
turn out to describe D2-branes on $SU(2)$ \cite{ars1}.

In this paper, we will develop a particularly simple formulation
of gauge theory on $S^2_N$, which makes a clear choice of the preferred
actions even in the nonabelian $U(n)$ case, and includes the topologically
nontrivial sectors in an extremely simple way. 
The starting point is the following observation: we can combine the 
generators $\la_i$ which are $N\times N$ matrices
into a single $2N \times 2N$  matrix by
\beq
C = \frac 12 + \sum_i \la_i \sigma^i.
\label{C-intro}
\eeq
We then observe the following property
\beq
C^2 = (\frac N2)^2,
\eeq
which follows from \eq{lasi2}. Hence the eigenvalues of $C$ are 
$\pm \frac N2$. To get the multiplicities,
note that $\la_i \sigma^i$ is an intertwiner of
$(2)\otimes(N) = (N-1) \oplus (N+1)$
(i.e. it is invariant under $SU(2)$), hence the
multiplicities are $N+1$ resp. $N-1$.

This simple observation leads to the idea that one should consider a 
matrix model for a hermitian matrix $C = \frac 12 + B_i \s^i$, and a potential
which has $\pm \frac N2$ as degenerate minima.
The fluctuations $B_i = \la_i +A_i$
around the above solution should correspond to 
the gauge fields $A_i$,
and the invariance under $C \to U^{-1} C U$ for a unitary
matrix $U$ should correspond to gauge transformations (and other symmetries).
Indeed this idea works. Before working it out,
let us briefly recall some basic facts about matrix models.

\section{Matrix Models and the Fuzzy Sphere}
\label{sec:3}

\subsection{A brief review of  single-matrix models}
\label{subsec:usual}

We briefly recall some basic facts about matrix models which
have found many applications in physics.
We refer the reader to \cite{m1,m2} for
excellent reviews and more references. 
Consider the matrix model of a single $N \times N$ hermitian matrix
$C$ with potential $V(C)$. 
The partition function of the model is defined by
\be \label{Z}
Z= \int dC e^{-\Tr V(C)} = \int \prod_{i=1}^N dc_i \, \Delta^2(c) 
\, e^{-\sum_i V(c_i)} 
\ee
where $c_i$ are the $N$ eigenvalues of the Hermitian matrix $C$.
Here
\be
\Delta (c) = \prod_{i < j} (c_i -c_j) 
\ee 
is the Vandermonde determinant, which is the Jacobian of the 
transformation $dC = \prod dc_i dU \Delta^2(c)$, and the integral over
the unitary matrices $U$ is trivial.
In the literature on matrix models, the
potential $V$ is usually  chosen to be of the form
\be \label{orig-V}
V (C) = \frac{N}{g^2} v(C), \quad v(C) = \sum_{k\geq 2} g_k \,  C^k
\ee
where $g_2 = 1$ and the couplings $g_k$ are kept 
fixed in the large $N$ limit.  

The reason why these matrix models are so useful is that
the models really only depend on the $N$ eigenvalues $c_i$, 
while the matrices have $N^2$ degrees of freedom.
This lead to the development of powerful methods
(e.g. steepest descent method, orthogonal polynomials, etc
\cite{zuber1,zuber2}) which can be used to analyze the models, and 
basically solve them explicitly in the large $N$ limit. 
For example, the saddle-point equation is given by
\be \label{sad}
\frac{1}{g^2} v'(c_i)
= \frac{2}{N} \sum_{j \neq i} \frac{1}{c_i -c_j}.
\ee 
The sum in the r.h.s.  is due to the Vandermonde determinant
in the measure and represents a repulsive potential among  the
eigenvalues. The $N$-dependence of the potential \eq{orig-V} ensures
that the repulsive effect is well balanced in the large $N$ limit.
Due to this repulsive force, the eigenvalues spread
evenly around the classical solution of the equation
of motion $c_i=0$. The distribution of the eigenvalues
\be
\rho(c) := \frac{1}{N} \sum_i \d(c-c_i)
\ee
becomes  continuous in the large $N$ limit and can be solved easily from
the equations \cite{zuber1}
\be
\frac{1}{g^2} v'(c) = 2 \int\kern-11.5pt-\kern7pt  
d c' \frac{\rho(c')}{c - c'},
 \quad  \quad \int dc \rho(c) =1.
\ee
There is much more to be said about these matrix models. In particular, the
distribution of eigenvalues (and correlation functions)
can be derived using e.g. the method of orthogonal polynomials, 
without relying on the saddle-point approximation.
However we will not need these techniques
due to the simplicity of the model considered here.

\subsection{A matrix model with degenerate minima}
\label{subsec:new}

\begin{figure}[htpb]
\begin{center}
\epsfxsize=3in
  \vspace{0.3in} 
   \epsfbox{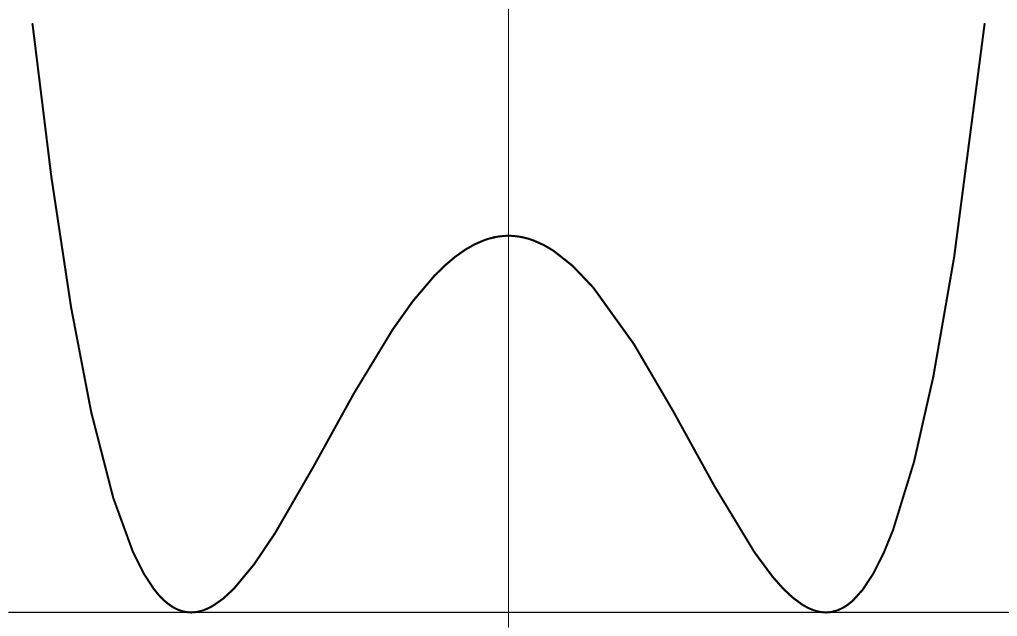}
\end{center}
 \caption{$V(C)$}
\label{fig:potential}
\end{figure}

In this paper, we will show that the fuzzy sphere arises as a vacuum
solution 
of another Hermitian matrix model, given by a potential
with a different scaling dependence in $N$. 
Consider the matrix model with action
\beq \label{V}
S = Tr V(C) =\frac{1}{g^2 N}  \Tr\left((C^2 -(\frac N2)^2)^2 \right)
   =\frac{1}{g^2 N} \Tr\left(\frac {N^4}{16} -\frac{N^2}{2} C^2 +  C^4\right) 
\eeq
where $C$ is a $2N \times 2N$ Hermitian matrix, and $g^2>0$ 
is kept fixed independent of $N$. 
The shape of the potential is sketched in Figure \ref{fig:potential}.
The following features distinguish it from the 
matrix models considered before:

\bit
\item The coefficient of the quadratic term is negative. This implies
that the distribution of eigenvalues are peaked 
at the minima $\pm \frac N2$ of the polynomial $V(C)$, 
rather than around the origin. 

\item
The specific form and the particular $N$ dependence of $V(C)$ 
is chosen such that the accumulation of eigenvalues at its minima will 
lead to the emergence of a (fuzzy) sphere $S^2_N$. 
The fluctuations will describe a gauge theory on $S^2_N$, 
as we will show below. Apart from the $N$ dependence, 
the relative coefficients between the $C^2$ and $C^4$ terms in $V$ can be 
adjusted to any (negative) number by a rescaling of $C$ and $g$. 
In general, different spaces may be generated for different potentials
$V(C)$, and the properties of this space such as 
symmetries are related to the details of this eigenvalue distribution.

\item If we expand $V(C)$ around one of its minima (consider 
$\frac N2$ to be specific, setting $C = \frac N2 + \mu_+$), then 
\be
V(\frac N2 + \mu_+) = \frac{N}{g^2 } (\mu_+^2 + \frac 2N \mu_+^3 + \frac
1{N^2}\; \mu_+^4)
\label{V-expand}
\ee
Now the coefficient of the quadratic term is positive, and
comparing with
the usual matrix models \eq{orig-V} we expect that the higher-order
terms will become irrelevant for large $N$, leading to a simple 
Gaussian distribution near each minimum.
\eit

Let us turn to the eigenvalue distribution in the large $N$
limit. The stationary points 
of our potential are given by the eigenvalues
\be
c= 0, \;\pm \frac{N}{2}
\ee
Since the action for $c=0$ is of order
$N^3$, it is quite obvious that this stationary point does not 
contribute for large $N$. 
Hence consider the eigenvalues distribution near
the minima $c = \pm \frac N2$ in the large $N$ limit. 
We introduce $c_i = \frac{N}{2} + \mu_i$ to facilitate
the analysis; the other case $c_i = -\frac{N}{2} + \mu_i$
is similar.  Performing a similar analysis as
\cite{zuber1}, the saddle point equation 
becomes
\be
\frac{1}{g^2}( 2\mu_i + \frac 6{N} \mu_i^2 + \frac 4{N^2} \mu_i^3) = 
\frac{1}{N} \sum_{j\neq i} \frac{2}{\mu_i - \mu_j}.
\ee
In the large $N$ limit this becomes again
\be
\frac{1}{g^2} \mu =  \int_{-a}^a d \mu' \frac{\rho(\mu')}{\mu - \mu'},
 \quad  \quad \int_{-a}^a d\mu \rho(\mu) =1
\ee
which gives the standard distribution for Gaussian matrices, 
\be
a = \sqrt{2g^2}\;, \qquad
\rho(\mu) = \frac 1{g \pi} \; \sqrt{2g^2-\mu^2}.
\label{e-dist}
\ee
In particular, the distribution is only nonzero for
$|\mu_i | < \sqrt{2}g$, and we find a {\em finite} spread of eigenvalues.
The reason is the explicit $N$ in front of the quadratic term in the
action. This means that only eigenvalues near $\pm\frac N2$ will contribute,
\be
\d \mu_i < \sqrt{2}g.
\label{fluct-bound}
\ee
Hence near each minimum $\pm \frac N2$, 
the action can be replaced by a Gaussian $\frac N{g^2}\sum_i (\mu^\pm_i)^2$
for large $N$,
since the higher-order terms are suppressed by $\frac 1N$.

\subsection{Matrix model vacua and emergence of the fuzzy sphere}
\label{sec:0-vacuum}

Let us resume the analysis of the particular model 
given by the polynomial \eq{V}. The
solutions to the classical equation of motion 
\be \label{eom2}
C (-N^2/4+ C^2) =0
\ee
are characterized by the multiplicities $n_+, n_-, n_0 $ of the 
eigenvalues $\pm \frac N2$ resp. $0$, which satisfy
$n_+ + n_- + n_0 = 2N$. We can assume that $n_0=0$ 
as discussed above, since each 
zero eigenvalue gives a contribution $\frac{N^3}{16 g^2}$
to the action and is highly suppressed.  Then 
the saddle points are characterized by the trace $\Tr(C)$, given by
\be
\Tr(C) =  N/2\; (n_+ -n_-).
\ee
Consider the vacuum with $n_+- n_- =2$, i.e. $n_+ =N+1, n_- =N-1$ . 
Using the $U(2N)$ invariance, one can put the
vacuum in the form 
\be \label{C0}
C = \frac{1}{2} +  \l_i \s^i
\ee
where $\l_i$ are precisely the $N \times N$ Hermitian matrices 
in \eq{lx}, \eq{FS-l} which describe the fuzzy sphere, 
and $\s^i$, $i=1,2,3$ are the Pauli matrices. 
In other words, we can find a unitary matrix $U$ such that
\be
\diag(\frac N2,....,-\frac N2) = U (\frac 12 + \l_i \s^i) U^{-1}
\label{U_0}
\ee
provided $n_+- n_- =2$, as explained in Section \ref{sec:basic}. 
The equations of motion
\eq{eom2} then take the form
\be \label{FS-soln}
\vare^{ij}_k \l_i \l_j = i \l_k,
 \quad \l_i \l^i =  \frac{N^2-1}{4}.
\ee
These are precisely the defining relation \eq{FS-l} of the fuzzy sphere
$S^2_N$ in terms of the dimensionless coordinates.
The radius has been set to unit here, since it 
can easily be reintroduced.

\subsection{Matrix fluctuations and gauge fields}
\label{subsec:fluct}

Now consider a general $2N \times 2N$ Hermitian matrix $C$,
\be
C = C_\a \s^\a 
  = (\frac{1}{2}+ \rho) \s^0  + B_i \s^i
\label{general-form}
\ee
where $\s^0 = \one$.
Plugging this into \eq{V}, we obtain
\bea \label{exp-action}
S = \Tr V(C) 
  = \frac {2}{g^2 N} && \!\!\!\!\!\!\!\!\Tr\Big((B_i B^i - \l_i \l^i)^2 +
  (B_i + i\vare_{ijk}B^j B^k)(B^i + i\vare^{irs}B_r B_s) \nn\\
 && + D_i\rho D^i\rho + N^2 \rho^2 + 2 \rho^3 + \rho^4 \nn\\ 
&& + 6(B_i B^i - \l_i \l^i)\rho(\rho+1)
 + 4 i \rho \vare_{ijk} (B^i B^j B^k - \l^i \l^j \l^k)\Big),
\eea
where
\be
D_i \rho := [B_i,\rho].
\ee
This starts to look like a field theoretic action on the fuzzy sphere
$S^2_N$.
Its interpretation is however obscured by the presence of $\rho$.
Comparing with \eq{C-intro}, we shall therefore impose the {\em constraint}
\beq
\rho =0, \quad  \mbox{i.e. }\; C_0 = \frac 12.
\label{constraint}
\eeq
This implies
\beq
\Tr(C) = N.
\eeq
Then the above action becomes 
\beq
S = \frac{2}{g^2 N} \Tr\Big((B_i B^i  - \frac{N^2-1}4)^2 
+ (B_i + i\vare_{ijk}B^j B^k)(B^i + i\vare^{irs}B_r B_s)\Big).
\label{B-action}
\eeq
This is one possible action for a gauge theory on
the fuzzy sphere, cp. 
\cite{madore,grossegauge,balagauge,klimcik,watamura,iso}.
The constraint breaks the original $SU(2N)$ symmetry down
to a smaller subgroup, which contains a $SU(N)$ gauge symmetry acting as
\beq
B_i \to U^{-1} B_i U. 
\eeq
We will see that this corresponds to the usual $U(1)$ local gauge
symmetry in the classical limit.
Of course, breaking the full $SU(2N)$ symmetry by the constraint 
\eq{constraint} is somewhat against the spirit of the matrix model, 
in particular the integration in the partition function 
cannot be carried out as easily as in \eq{Z} any more.
This is to be expected, since \eq{B-action} is a matrix model with 
3 interacting matrices. Nevertheless, this approach will
allow us to carry out the path integral with some more effort. 

It is easy to understand the dominant configurations for the
action \eq{B-action}. The term 
$(B_i + i\vare_{ijk}B^j B^k)^2$ implies that the 
$B_i$ approximately generate a representation of 
$su(2)$, and the other term implies that $B_i B^i \approx (N^2-1)/4$,
which corresponds to the Casimir of the approximate $su(2)$
representation.
Hence  one could interpret \eq{B-action} as a theory of ``fluctuating
representations'' of $su(2)$, and
the dominant configurations
will be approximately $N$-dimensional irreps of $su(2)$.
This is an important difference to other
possible actions without the term $(B_i B^i  - \frac{N^2-1}4)^2$, 
such as in \cite{ars2}: 
there, reducible ``block''-solutions with blocks of arbitrary size 
are allowed, while in \eq{B-action} they are suppressed. As we will
see, this is crucial for the physical interpretation, and only 
\eq{B-action} reduces to an ordinary Yang-Mills
theory on $S^2$ in the large $N$ limit. 

Consider next the equations of motion, 
\be
[B^i, B_j B^j  - \frac{N^2 - 1}4]_+  +  (B+i\vare BB)_i 
   + i \vare^{ijk} [B_j, (B+i\vare BB)_k]  = 0.
\label{eom}
\ee
The ``vacuum'' solution is
\be
B_i = \l_i,
\label{B-la}
\ee
up to gauge transformation.
In fact then $S=0$, and this is the unique solution with $S=0$
up to $SU(N)$ gauge invariance, because both
$B+i\vare BB =0$ and $B_i B^i = \frac{N^2-1}4$ must hold. This  
means that $B_i$ is a representation of $su(2)$ with fixed Casimir.
If we now expand
\beq
B_i = \l_i + \cA_i,
\label{B-split}
\eeq
then
\be
 B_i B^i - \l_i \l^i =
\l_i \cA^i + \cA_i \l^i + \cA_i \cA^i
\ee
and
\be 
B^i + i\vare^{ikl}B_k B_l = \frac{1}{2}\vare^{ikl} F_{kl}, \quad
F_{kl}:= i [\l_k,\cA_l] - i [\l_l,\cA_k] + i [\cA_k,\cA_l] + \vare_{klm} \cA^m.
\label{F-def}
\ee
Notice that the kinetic terms in $F_{kl}$ arise 
automatically due to the shift \eq{B-split}.
The $SU(N)$ gauge symmetry acts on $\cA_i$ as
\be
\cA_i \to  U^{-1} \cA_i U +  U^{-1} [\la_i, U]
\ee
which for $U = \exp(i h(x))$ and $N \to \infty$ 
becomes the usual (abelian) gauge
transformation for a gauge field.

However, the gauge field $\cA_i$ has 3 components, which does  not
seem to match with the degrees of freedom in a 2-dimensional gauge
theory. To understand this, we should translate the action
into  conventional field theory language,
which can be done for large $N$.
One can then decompose the field $\cA_i$ into a tangential component
$A_i$ and a radial component $\varphi$  as follows\footnote{this
  decomposition as defined here is gauge-invariant only in the large $N$
  limit. From that point of view  
$\tilde \varphi:=  B_i B^i - \l_i\l^i$ would be a nicer radial field,
however this leads to a nontrivial Jacobian in the path
integral, which makes the decoupling argument below more subtle.}:
\be
\cA_i = \frac{4}{N^2-1}\; \l_i\,\varphi + A_i\; 
       \approx \frac{2}{N} x_i\,\varphi + A_i
\ee
where $\varphi$ and $A_i$ are defined such that
\bea 
\l^i A_i  &=& 0, \label{t2}\\
\varphi &:=& \l^i \cA_i. \label{t1}
\eea
Then
\be
 B_i B^i - \l_i \l^i = 2 \varphi 
   -[ \l^i, A_i] +A_i^2 \; +\frac 1N \; T(\varphi, A).
\label{BB-explicit}
\ee
Here $T(\varphi, A)$ stands for  functions of $\varphi$ and $A_i$ 
which are suppressed by  $\frac 1N$. 
Similarly, all terms involving
$\varphi$ in the ``field strength'' $F_{kl}$ \eq{F-def} are 
suppressed\footnote{assuming that $\varphi$ and $A_i$ is ``smooth'',
so that $[\varphi,\l_i]$ is finite. This is justified by
\eq{phi-replace} and the kinetic terms in the action.} by
$\frac 1N$.
Therefore the only term involving  $\varphi$ which contributes for
large $N$
is the square of \eq{BB-explicit}. We can now simply integrate out $\varphi$
(i.e. consider it an auxiliary variable), 
replacing it by 
\be
\varphi = \frac 12([ \l^i, A_i] - A_i^2)
\label{phi-replace}
\ee
for large $N$, which is smooth (i.e. high angular momenta are 
suppressed, assuming the $A_i$ are smooth). 
Hence all terms in $F_{kl}$ containing $\varphi$
can be omitted for large $N$, being suppressed by $\frac 1N$.
Another way of arriving at this conclusion is to use the field
$\phi:=\frac 1N \varphi$, which has a large mass of order $N$. 
At any rate, we can now write 
\be 
\label{nc-field}
F_{kl}= i [\l_k,A_l] - i [\l_l,A_k] + i [A_k,A_l] + \vare_{klm} A^m
\ee
for large $N$, involving the tangential gauge field only.
As a consequence of \eq{t2} and \eq{FS-l}, 
$F_{kl}$ is ``tangential'' 
\be \label{t3}
x^k F_{kl} =o(1/N)
\ee
for large $N$, and it becomes 
the field strength of an abelian gauge theory on $S^2$. 

To summarize, we found that the radial fluctuations
$\varphi$ decouple in the large $N$ limit, and 
\eq{B-action} reduces to a $U(1)$ Yang-Mills 
theory on a unit sphere  with action
\be \label{YM-action}
S  = \frac{1}{g^2}\int F_{mn} F^{mn}.
\ee 
Here 
\be
F_{kl} = i J_k A_l - i  J_l A_k + \vare_{klm} A^m
\ee
is the  field strength for the $U(1)$ gauge potential $A_i$. 
The fields are tangential in the sense \cite{qFSI}
\be \label{t4}
x^i A_i =0, \quad x^i F_{ij} =0.
\ee
This is a description of the 2d gauge theory  in terms of a
3-component gauge field $A_i$ subject to the tangential constraint
\eq{t4}.  This formulation is manifestly invariant under $SO(3)$
rotations. To put it in a more familiar form, let us assume 
that we are sitting on the north pole. 
Then only  $A_1,A_2$ and $F_{12}$ survive the constraints, and 
\be
i J_1 = - \del_2, \quad i J_2 = \del_1.
\ee 
Our  gauge theory can then be identified with a gauge theory with
only tangential gauge fields $A^{(cl)}_i$, $i=1,2$, 
whose field strength takes the usual form 
\be
F^{(cl)}_{12} = \del_1 A^{(cl)}_2 - \del_2 A^{(cl)}_1
\ee
if we identify 
\be \label{iden}
A^{(cl)}_1 = -A_2, \quad A^{(cl)}_2 = A_1.
\ee
In coordinate independent form, this is
\be
\vec A^{(cl)} = \vec r \times \vec A
\label{iden-gen}
\ee
where $\vec r$ is the radial unit vector. 
This identification will be useful in the next subsection.

Since the volume of the 
gauge group is finite here, we do not have to fix the $SU(N)$ gauge 
using e.g. the Faddeev-Popov method. Instead we can keep the integral
over all configurations. 
Indeed, working with $B_i$ or even $C$ seems much easier,
and  makes all the symmetries manifest. 
The beauty of our formulation is that it allows to apply the 
powerful methods of random matrix theory, 
after suitable modifications. 
One can hope that this will lead to new methods
for studying gauge theories.

\paragraph{Alternative version of constraint.}

It is important to realize that the matrix model $\Tr V(C)$ describes a 
Yang-Mills theory only if we impose the constraint \eq{constraint}, so that
the last terms in \eq{exp-action} vanishes. Without that constraint, 
$4 i \rho \vare_{ijk} (B^i B^j B^k - \l^i \l^j \l^k)$
contains a term $\propto N F \rho$, which after integrating out $\rho$
cancels the YM term $\Tr FF$. However, there
is another possibility, namely to consider
\beq
S' = \Tr(V(C)) - N \Tr(C_0 - \frac 12)^2.
\eeq
The last terms of course implies the $C_0 = \frac 12$ constraint in the
large $N$ limit. This leads to an additional term $-\Tr N \rho^2$ in the
action \eq{exp-action}, which allows to integrate out $\rho$ for large $N$
leaving a rescaled YM term $\propto \Tr FF$. 
The path integral can then 
be carried out in the same way as we will do below. Basically this seems 
to be a matter of taste, and we chose to impose $\rho =0$ directly.

\subsection{Monopole sectors}
\label{subsec:monopole}

If we claim to have a fuzzy version of  Yang-Mills theory on the 2-sphere,
we should be able to recover the monopole sectors as well. 
They are  not hard to find here:
as discussed above, the dominant
contributions for the action \eq{B-action} are ``approximate'' representations
of $su(2)$ with Casimir  $B_i B^i \approx \frac{N^2-1}4$.
This suggests to consider irreps of dimension $M$ slightly different from $N$.
They indeed turn out to describe monopoles\footnote{The basic
idea that fuzzy spheres of different size correspond to monopoles
was also proposed in \cite{nair}, without calculating the gauge field.}.

Hence consider the 
same action \eq{V}, but for  matrices of different size. Let
\be
C^{(M)} = \frac 12 + B^{(M)}_i \sigma^i
\ee
be a $2M \times 2M$ matrix
with
\beq
M = N-m, \quad m \in \Z.
\eeq
This implies in particular
\beq
Tr(C) = M,
\eeq
which again picks out the sector $n_+- n_- =2$.
One can easily see that the equation of motion \eq{eom} 
resp. $Tr(V'(C) \d C)=0$ 
has solutions of the form $B^{(M)}_i = \a_m \la^{(M)}_i$
if 
\beq
3(\a_m-1)^2 +\a_m^2 M^2 -N^2 =0, 
\eeq
which gives 
\beq
\a_m = 1+\frac mN \qquad \mbox{ for } N \gg m.
\label{alpha}
\eeq
Hence we found the new solution
\be
C^{(M)} = \frac 12 + \a_m \la^{(M)}_i\sigma^i.
\label{mono-M}
\ee
Then
\be
(F^{(M)})_i = i\vare_i^{jk} (B^{(M)})_j  (B^{(M)})_k + (B^{(M)})_i
 \; \to \; \frac m2\; x_i
\label{F-monopole}
\ee
and 
\be
B_{(M)}\cdot B_{(M)} -  \frac{N^2-1}4
    \;\to\; 0 
\ee
for $N \to \infty$.
In particular, the field strength is tangential in the sense \eq{t3},
with $|F| = \frac m2$. This is just like 
the field strength of a monopole of charge $m$,
suggesting that $m$ is the monopole charge. The action is 
\beq
S(C^{(M)}) = \frac{m^2}{2g^2}
\label{mono-action}
\eeq
for large $N$. 
We will show that this interpretation as a monopole is correct by 
writing $(B^{(M)})_i$ as an excitation over the 
fuzzy sphere solution $B_i = \l_i^{(N)}$. The corresponding gauge field
will take the usual form of Dirac monopole
of charge $m$ in the large $N$ limit. To see this, write this solution
as a block matrix
\be
C = \frac 12 + \left(\begin{array}{cc} 
\a_m \la^{(M)}_i\; \sigma^i & 0
\\ 0 & 0\end{array}\right) = \frac 12 + \la_i^{(N)} \sigma^i
     + A_i \sigma^i,
\label{monopole-matrix}
\ee
where $A_i$ is interpreted as gauge field. Hence
\beq
A_i = \a_m \la^{(M)}_i - \la_i^{(N)}.
\eeq
Using the representation \eq{reps} for the $\l_i = \l_i^{(N)}$, one obtains the
following non-vanishing matrix elements
\be \label{k2}
(A_3)_{kk} = \left\{\begin{array}{ll}
-(\l_3)_{kk} \left( 1- \a_m (1-\frac{m}{N+1-2k}) \;\right)
&  1\leq k\leq N-m,\cr 
-(\l_3)_{kk}  , & k > N-m,
\end{array}
\right.
\ee
\be \label{k3} 
(A_+)_{k,k+1} = \left\{ \begin{array}{ll}
-(\l_+)_{k,k+1} \left(1-
\a_m\sqrt{1-\frac{m}{N-k}}\;\right), & 1\leq k\leq N-m, \cr
-(\l_+)_{k,k+1}  , & k > N-m
\end{array}
\right.
\ee
and $A_-$ is obtained by transposition. 
This can be translated into functions on the 
fuzzy sphere via \eq{lx}, which takes the form 
\be \label{lx1}
x_i = \frac{2}{N} \l_i
\ee
in the large $N$ limit. We also note that
$k$ is related to the ``height'' $x_3$ through 
\be
k = \frac{N+1}{2} - (\l_3)_{kk}.
\ee
The quantities in \eq{k2}-\eq{k3} have a smooth limit for large $N$, 
except at the finite set of ``points'' $N-m+1<k< N$ 
(located at the south pole) where $A_i$ develops
a singularity. This singularity corresponds to the Dirac string. 
In the patch covered by $1\leq k\leq N-m$, which represents
the sphere without the south pole, we obtain 
\be \label{A3}
A_3 = -\frac{m}{2} (1-x_3),
\ee
\be \label{A+}
A_+ = - \frac{m x_+}{2 (1+x_3)} + \frac{m x_+}{2},
\ee
\be \label{A-}
A_- = - \frac{m x_-}{2 (1+x_3)} + \frac{m x_-}{2},
\ee
in the large $N$ limit (recall that $R=1$ throughout).
It is easy to check that $A_i$ satisfy the constraint \eq{t4}. 
This looks almost but not quite right; however, recall that we must 
use the identification \eq{iden-gen} 
to find the corresponding classical gauge field $A^{(cl)}_i$. 
This comes indeed out as
\be
\vec A^{(cl)} = \vec r \times \vec A = \frac m2 \frac 1{1+x_3}\;
\left(\begin{array}{c} x_2\\-x_1\\0\end{array}\right)
\ee
or $A^{(cl)} = \frac m2 \frac 1{1+x_3}\;(y dx - x dy)$, 
which is precisely the (tangential) gauge  field of a Dirac monopole
of charge $m$ on the sphere. 
The field strength was already calculated in 
\eq{F-monopole},  and is constant with the correct quantization 
\beq
 \int F \equiv  \int F_i x^i = 4\pi\frac m2
\eeq
Notice that $F$ is constant in spite of (or rather because of)
the ``non-classical'' term $[A_i, A_j]$ in the definition of $F$.
The same calculation applies for $m<0$, hence we get both negative and
positive monopole charge as it should be. 
The singularity at the south pole can of course be moved around using
suitable $SU(M)$ gauge transformations.
At finite $N$ resp. $M$, this configuration 
should therefore be interpreted as a fuzzy monopole. 

This point of view considering the monopole sectors as matrices
of different size is quite compatible with the treatment of 
nontrivial topological sectors in \cite{grosse-topology},
where sections in nontrivial
bundles are represented by $N \times M$ matrices. Clearly
our gauge fields $B_i$ of the appropriate size
can act on these from the left resp. right, and one can define 
covariant derivatives in this way. This will be elaborated elsewhere.

\paragraph{More careful embedding of monopole sectors.}

We should address a somewhat unsatisfactory aspect of 
the above treatment of the monopole sectors: Formally
we have been considering distinct matrices $C^{(M)}$ for different
$M$, but more properly they should all be considered 
as block-matrices embedded
\beq
... \hookrightarrow Mat(N-1)
  \hookrightarrow Mat(N) \hookrightarrow Mat(N+1)  \hookrightarrow ...
\label{embedding}
\eeq 
as in \eq{monopole-matrix}, cp. also \cite{grosse-susy}.
Then there is a small problem with
\eq{monopole-matrix}: the $\frac 12$ in the 
lower-right block in \eq{monopole-matrix} must be there in order to satisfy 
the constraint $\rho = 0$. However the eigenvalues of this small block
are far from $\pm \frac N2$. Therefore this type of block-matrix 
configuration would strictly speaking
be highly suppressed by the action, because the
Dirac-string contributes $o(N^3)$ to the action.
One way to cure this 
problem is to replace the action by
\beq
S' = \Tr (\frac{(C-\frac 12)^2}{N^2} V(C)),
\label{Vprime}
\eeq
which now has $\pm \frac N2, \frac 12$ as degenerate minima, 
and all fluctuations are Gaussian as in \eq{V-expand}.
Here $C \in Mat(M)$ should be a matrix of fixed size $M$ 
which is large enough to accommodate all relevant solutions,
i.e. $N \ll M \ll 2N$.
Now the block-matrix \eq{monopole-matrix} is really 
a solution of the equation of motion, with action $\frac {m^2}{2g^2}$.
In terms of the $B_i$ fields, we note that 
$(C-\frac 12)^2 = B_i B^i + i\vare^{ikl}B_k B_l \s^k$, so that the action
differs from \eq{B-action} by terms of the form
$\Tr((\frac{\varphi}{N^2} + \frac iN x_i \s^i)V(C))$
which are suppressed for large $N$.
The crucial difference is that the Dirac-string now has vanishing action.
Therefore all monopole configurations can be obtained
as distinct solutions in one single configuration
space $Mat(M)$. This is conceptually very 
appealing, because it shows that the nontrivial topological 
sectors arise here automatically as different solutions
for the same action. This is even simpler than in the classical
case: there is no need to introduce nontrivial 
principal bundles, they just
come out. However the calculation of the partition function
below would be somewhat more complicated for the action \eq{Vprime}.
Since the path integral is the main
focus of this paper, we shall not pursue this point of view
here, and consider the monopole configuration as 
truly ``distinct'' sectors for simplicity, as classically.

\section{Nonabelian case: $U(n)$ Yang-Mills theory}
\label{sec:nonabelian}

Now consider the same matrix model $S = \Tr V(C)$ as in \eq{V}, but for
larger matrices of size $2M \times 2M$ with
\beq
M = nN-m, \qquad \Tr(C) = M.
\eeq
The last constraint implies that the multiplicities of the (dominant) 
eigenvalue distributions of $C$ are now $n_+ - n_- = 2n$. 
We will see that this leads to a non-abelian $U(n)$ Yang-Mills theory. 

First we should find the ground state. For $M=nN$, 
the absolute minima of the action $S= \Tr V$ are now
given by any matrix $C$ with $n_+ = M+n$ eigenvalues $+\frac N2$ and 
$n_- = M-n$ eigenvalues $-\frac N2$. In a suitable basis, $C$ takes the form 
\be \label{ground-C-nonabel}
C = (\frac 12 + \la_i^{(N)} \sigma^i)\; {\bf 1}_{n\times n},
\ee 
which is a block matrix consisting of $k$ blocks of the solutions
$(\frac 12 + \la_i^{(N)} \sigma^i)$ of Section \ref{sec:0-vacuum}.
The action is then zero, and clearly all other saddle points 
have a positive action. 
In general, we can write again
any $2M \times 2M$ matrix $C$ in the form 
\be \label{fluc-C-nonabel-1}
C = (\frac 12 +\rho) + B_i \sigma^i
\ee 
where the $B_i$ and $\rho$ are now $M \times M$ matrices.
In order to obtain a Yang-Mills gauge theory, we shall impose 
again the constraint 
\beq
\rho = 0,
\eeq
so that the action \eq{V} reduces to
\beq
S = \frac {2}{g^2 N} \Tr\Big((B_i B^i  - \frac{N^2-1}4)^2 
+ (B_i + i\vare_{ijk}B^j B^k)(B^i + i\vare^{irs}B_r B_s)\Big),
\label{B-action-again}
\eeq
which has the same form as \eq{B-action} but for different 
size $M$ of the matrices.
It is invariant under the gauge transformations $B_i \to U^{-1} B_i U$
for $U \in U(M)$. To understand its meaning, we write 
the fluctuations of $B$ resp. $C$  in the form
\beq    \label{fluc-C-nonabel-2}
    B_i = B_{i,\a} t^\a = \l_i^{(N)}\; t^0  + \cA_i
\eeq 
where $B_i$ and $\cA_i$ carry a $u(n)$ index,
\be 
\cA_i =  \cA_{i,0}\; t^0 + \cA_{i,a}\; t^a.
\ee
Here $t_a$ denote the Gell-Mann matrices of $su(n)$, which satisfy
\be
t_a t_b = \frac 1n g_{ab} +  \frac 12 (d_{ab}{}^c  +  if_{ab}{}^c) t_c
\ee
and $t_0 = \bf{1}$ is the $n\times n$ unit matrix. 
The rest of the analysis of Section \ref{subsec:fluct}
goes essentially through. In particular,
we can split the gauge fields again into
tangential and radial components
\be
\cA_i = \frac {4 \l_i}{N^2-1} \,\varphi + A_i.
\ee
We suppress here the $u(n)$ labels of $\varphi$ and $A_i$,
which are defined by
\beq
\varphi :=  \l^i \cA_i, \qquad \l^i A_i  = 0.  
\eeq
Then \eq{BB-explicit}
implies as before that all components of $\varphi$
decouple and can be integrated out. It remains
\be
B_i =  \l_i +  A_i = \l_i + A_{i,0} t^0 + A_{i,a} t^a
\ee
involving only the tangential components of $\cA_i$, and 
in the large $N$ limit
we obtain a theory  with action 
\be \label{YM-action-nonabel}
S = \Tr V = \frac{1}{g^2}\int F_{mn} F^{mn}
\ee 
where again
\be
B^i + i\vare^{ikl}B_k B_l = \frac{1}{2}\vare^{ikl} F_{kl}, \quad
F_{kl} = i [\l_k,A_l] - i [\l_l,A_k] + i [A_k,A_l] + \vare_{klm} A^m
\ee
is tangential 
\be 
x^k F_{kl}  = o(1/N).
\ee
Spelling out the $u(n)$ structure explicitly and omitting terms 
which vanish for large $N$, this action becomes
\be \label{YM-action-nonabel2}
S = \Tr V = \frac{1}{g^2}\int (F_{mn,0} F^{mn,0} + F_{mn,a} F^{mn,a}),
\ee 
where 
\bea
F_{kl,0} &=& i J_k A_{l,0} - i  J_l A_{k,0} + \vare_{klm} A_{m,0}, \nn\\
F_{kl,a} &=&  i J_k A_{l,a} - i  J_l A_{k,a} + i A_{k,b} A_{l,c} f_a^{bc} 
      + \vare_{klm} A_{m,a}.
\eea
This is the action of a $U(n)$ Yang-Mills theory on the sphere. 
Recall that the only difference
to the abelian case in Section \ref{subsec:fluct} 
is the size $M \approx nN$ of the matrices.

\paragraph{Saddle points.}

The remaining saddle-points can now be found from 
the equation of motion \eq{eom} as in the previous section. 
Clearly any (reducible, in general) 
representation of $su(2)$ with a suitable normalization as in \eq{mono-M}
will give a
solution. Therefore the (dominant) saddle-points are given by
the block-matrices 
\be
C^{(m_1, ..., m_n)} = \left(\begin{array}{cccc}
  C^{(M_1)} & 0 & \dots & 0 \\
  0 & C^{(M_2)} & \dots & 0 \\
  \vdots & \vdots & \ddots & \vdots \\
  0 & 0 &  \dots & C^{(M_n)} \end{array} \right),
\label{saddle-C}
\ee
where each block has the form \eq{mono-M} with size
\beq
M_i = N - m_i, \qquad m_i \in \Z,
\eeq
such that $m_1 + ... + m_n = m$. Their action is 
\beq
S(C^{(m_1, ..., m_n)}) = \frac1{2g^2}\; \sum_i m_i^2
\label{mono-action-2}
\eeq
for large $N$, hence configurations with
large $|m_i|$ are suppressed.

We can now write the saddle-points \eq{saddle-C} as fluctuations 
around the ground state, as in Section \ref{subsec:monopole}. 
After arranging the blocks appropriately
(by a gauge transformation), they take the form
\beq
A_i = \left(\begin{array}{cccc}m_1 A_i & 0 & ... & 0 \\
                               0 & m_2 A_i & ... & 0 \\
                               \vdots & \vdots & \ddots & \vdots \\
                               0 & 0 & ... & m_n A_i \end{array}\right)
\eeq
(for large $N$) with $m_1 + ... + m_n = m$,
where $A_i$ is the basic abelian monopole field found in Section 
\ref{subsec:monopole}. Hence 
the sectors with $m_i \neq 0$ correspond to 
nontrivial $U(n)$ gauge field configurations, 
which are precisely the ``instantons''\footnote{we refer to any
critical point of the YM action as instanton, as is customary in the 
related literature. In general they are unstable.}
of the $U(n)$ YM theory found in \cite{witten2,gross}.

\section{The path integral}
\label{sec:pathintegral}

The quantization of
gauge theory on the 2-sphere has been studied extensively, using a
variety of methods including lattice formulations and 
a generalization of the 
Duistermaat-Heckmann localization theorem, see e.g.
\cite{migdal,rusakov,witten2,gross,witten1,poly,Douglas-kazakov,kazakov}. 
In particular, the partition function and  correlation functions
of Wilson loops have been calculated. It is therefore
natural to ask whether such calculations can also be done on the fuzzy
sphere. Some of the known methods, in particular the localization
theorem, might well be applicable as shown in 
\cite{szabo} for the case of a torus. 
However this method is rather 
indirect, and we want to calculate the path integral directly taking
advantage of the above formulation as matrix model.

One of the nice features of the fuzzy sphere is the fact 
that all path integrals are finite, simply because there 
are only finitely many degrees of freedom. However, this does not 
necessarily make them easy to evaluate: e.g. for scalar fields,
one is forced to resort to perturbation theory (see e.g. \cite{fuzzyloop}), 
which is even more complicated than in the classical case.

The main advantage of our matrix formulation of gauge theory is 
that it allows to explicitly carry out the path integral. This 
provides a truly 
new approach to gauge theory, since this formulation is possible only
in the noncommutative case. Without the constraint
$\rho = 0$, the integration would even
be ``trivial'' as in Section \ref{subsec:usual}, but this 
does not describe a  YM theory on the sphere. 
We will now show how this constraint can be handled using the known 
matrix model technology, and calculate the partition function directly 
by integrating over the gauge fields for large $N$. We will
recover the known result \cite{migdal,rusakov} for 
the partition function of $U(n)$ YM theory on the sphere for $N \to \infty$.
While the explicit calculation in Appendix B
may seem a bit involved for our present application, 
one can hope that the idea will be useful
in less ``trivial'' cases as well.

We want to quantize the gauge theory with action \eq{B-action-again} 
by integrating over the
$M \times M$ matrices $B_i$. This will be done by 
integrating over the $2M \times 2M$ matrices $C$ in the action \eq{V},
imposing the constraint \eq{constraint}. 
We will not attempt here to calculate the full generating functional 
for the gauge field, only the partition function
\bea
Z &=& \int d B_i\; \exp(-S(B))) \nn\\
 &=& \int d C\;  \d(C_0 - \frac 12)\; \exp(-\Tr V(C)) \nn\\
 &=& \int d\L_i \D^2(\L_i) \exp(-\Tr V(\L)) 
     \int dU \delta((U^{-1}\L U)_0 - \frac 12)
\label{Z-noJ}
\eea
where $dU$ is the integral over $2M\times 2M$ unitary matrices,
and $C = U^{-1}\L U$. 
Here $\delta(C_0 - \frac 12)$ is a product over $M^2$
delta functions, which 
can be calculated as follows: define
\be  
J = \left(\begin{array}{cc} K & 0 \\ 0 &  K \end{array}\right)
 = K\; \s^0
\label{J}
\ee
where $K$ is a $M\times M$ matrix. Then
\be
\delta((U^{-1}C U)_0 - \frac 12) = \int dK 
\exp(i \Tr(U^{-1} (C -\frac 12) U J)).
\ee
By  gauge invariance, the r.h.s. depends on the eigenvalues $\L_i$ of $C$
only. 
Hence
\beqa
Z  &=& \int dK  \int d\L_i \D^2(\L_i) \exp(-\Tr V(\L)) 
    \int dU \exp(i \Tr(U^{-1}\L U J - \frac 12 J)) \nn\\
 &=&   \int dK\;  Z[J]\; e^{-\frac i2 \Tr J}
\label{Z-J}
\eeqa
where
\beq
Z[J] := \int d C\;  \exp(-\Tr V(C) + i \Tr (C J))
\eeq
depends only on the eigenvalues $J_i$ of $J$.
Diagonalizing $K = V^{-1} k V$, we get 
\beq
Z  = \int d k_i  \D^2(k) \int d\L_i \D^2(\L_i) \exp(-\Tr V(\L)) 
    \int dU \exp(i \Tr(U^{-1}(\L-\frac 12) U J)) 
\label{Z-J2}
\eeq
where $\int dV$ was absorbed in $\int dU$. 
The main step is now to carry out the integral over $\int dU$, 
which can be done using the Itzykson-Zuber-Harish-Chandra formula 
\cite{IZ2,harish},
\be
\int dU \exp(i \Tr(U^{-1} C U J)) = const 
  \frac{\det(e^{i \L_i J_j})}{\D(\L_i) \D(J_j)}.
\label{U-int}
\ee
This depends only on the eigenvalues of $J$ and $C$, 
with  Vandermonde-determinants $\D(\L_i)$ and $\D(J_i)$. 
Note that the Vandermonde-determinants 
are totally antisymmetric, and so is $\det(e^{i \L_i J_j})$.
Therefore this expression is manifestly symmetric in both $\L$ and $J$.

In this step we have reduced the number of integrals from $M^2$ to 
$2M$. This means basically that the integral over fields on 
$S^2_N$ is reduced to the integral over functions in one variable. 
This is a huge step, just like in the usual matrix models. The constraint
however forces us to evaluate in addition the integral over $k_i$, 
which is quite complicated due to the rapid oscillations in 
$\det(e^{i \L_i J_j})$; recall that $\L_i \approx \pm \frac N2$. 
Nevertheless, it is shown in Appendix B how the integrals over
$\L_i$ and $k_i$ can be evaluated for large $N$, with the result 
\beq
Z_{m} =  \sum_{m_1 + ... + m_n = m}\; 
     \int_{-\infty}^{\infty} d\k_1... d \k_n\;\D^2(\k) \;  e^{i \k_i m_i}\;
    \exp(-\frac{g^2}{2}\sum \k_i^2).
\label{Z-n-nice}
\eeq
for matrices of size $M = nN-m$ (omitting overall constants).
This form of $Z$ was found in \cite{poly} for a $U(n)$ Yang-Mills
theory on the ordinary 2-sphere, apart from the constraint $\sum m_i =m$ 
which will be removed soon. 
It can be rewritten in the ``localized'' form as a weighted sum of
saddle-point contributions, as advocated by Witten \cite{witten2}. 
This can be seen as follows:
\beqa
Z_{m} &=& \sum_{m_1 + ... + m_n = m}\; \D^2(-i\frac{\del}{\del m_i}) \;
    \int d \k_i\  e^{i \k_i m_i}\;
    \exp(-\frac{g^2}{2}\sum \k_i^2)  \nn\\
 &\propto& \sum_{m_1 + ... + m_n = m}\; \D^2(\frac{\del}{\del m_i}) \;
   \exp(-\frac{1}{2g^2}\sum m_i^2) \nn\\
 &\propto& \sum_{m_1 + ... + m_n = m}\; \D(\frac{\del}{\del m_i}) \;
    \D(m_i)\; \exp(-\frac{1}{2g^2}\sum m_i^2) \nn\\
 &=& \sum_{m_1 + ... + m_n = m}\;  P(m_i,g) \; \exp(-\frac{1}{2g^2}\sum m_i^2).
\label{Z-9}
\eeqa
Here $P(m_i,g)$ is a totally symmetric polynomial in the $m_i$.
The last exponential is precisely
the action \eq{mono-action-2} for the saddle-point 
$(m_1, ..., m_n)$ as discussed in Section \ref{sec:nonabelian}, 
which is weighted by the polynomial $P(m_i,g)$ 
(e.g. for $n=2$, one finds $P(m_i,g) = (m_1-m_2)^2 - 2g^2$).
This shows that the ``localization'' \cite{witten2}
also holds in the noncommutative case for gauge theory on the 
fuzzy sphere, at least in the large $N$ limit. 
However we did not use any localization theorem here, 
it comes out by an explicit computation of the 
purely bosonic path integral, without having to introduce
auxiliary fermionic fields as in \cite{witten2}.
Note in particular that 
we did not do any gauge-fixing, which is not necessary here
because the volume of the gauge group $U(M)$ is finite.

To include all monopole configurations, 
we should sum over matrices of different sizes $M = nN -m$ 
as explained in Section \ref{sec:nonabelian},
keeping $V(C)$ constant. Then $m$ is the $U(1)$ monopole
charge. Hence the full partition function is obtained by
summing\footnote{the relative weights of $Z_m$ for different $m$ is strictly
speaking not determined here. However, it could be calculated using
the embedding \eq{embedding} as explained in Section \ref{subsec:monopole}}
over all $Z_m$, 
\beq
Z = \sum_m Z_m = \sum_{m_1, ...,m_n = -\infty}^{\infty} \; 
     \int d \k_i\;\D^2(\k) \;   e^{i \k_i m_i}\;
    \exp(-\frac{g^2}{2}\sum \k_i^2).
\label{Z-full}
\eeq
One can now  perform a Poisson-resummation as in \cite{poly},
\beq
\sum_m f(m) = \sum_p \tilde f(2\pi p)
\eeq
where $\tilde f(p) = \int \frac{dx}{2\pi} f(x) e^{-ipx}$. This gives 
\beq
Z = \sum_{p_1, ..., p_n \in \Z}\; \D^2(p) \; 
    \exp(-2\pi^2 g^2 \sum_i p_i^2).
\eeq
This is the partition function of a $U(n)$ Yang-Mills
theory on the ordinary 2-sphere. As shown in \cite{poly},
this is equivalent to the form found in \cite{migdal,rusakov}
\beq
Z = \sum_{R}\; (d_R)^2 \; 
    \exp(-4\pi^2 g^2 C_{2R}),
\eeq
where the sum is over all representations of $U(n)$ and $d_R$ is the
dimension of the representation and $C_{2R}$ the quadratic casimir.
Hence the limit $N \to \infty$ of the partition function
for $U(n)$ YM on the fuzzy sphere is well-defined, and 
reproduces the result for YM on the classical sphere. 
This strongly suggests that the same holds for the full 
YM theory on the fuzzy sphere, and that
there is nothing like UV/IR mixing for pure gauge theory on $S^2_N$.
This is unlike the case of a scalar field,
which exhibits a ``non-commutative anomaly'' \cite{fuzzyloop} 
related to UV/IR  mixing.

\section{Remarks on symmetries and correlation functions.}
\label{sec:symmcorr}

Let us try to understand the symmetries of our model in more detail.
Recall that  2-dimensional Yang-Mills theory is invariant under
area-preserving diffeomorphisms (APD's),
because the field strength $F = F_{ij} dx^i dx^j = f \omega$ 
has only one component, and $\int F\star F = \int f^2\; \omega$
is invariant under APD's. Here $\omega$ denotes the volume form.

It is easy to understand the quantization of 
APD's on the fuzzy sphere\footnote{I want to thank S. Rajeev
for explaining this to me}.
The fuzzy sphere arises as quantization of the 
Poisson structure $\{x_i, x_j\} = \vare_{ijk} x_k$,
which corresponds to the (canonical) symplectic 
form $\omega = x^i dx^j dx^k \vare_{ijk}$  on $S^2$ which coincides
with the volume form.
Hence any function $f(x)$ on the sphere defines a 
Hamiltonian flow, which preserves $\omega$. 
It is therefore an area-preserving diffeomorphism.
Explicitly, the vector field $X_f$ generating the APD with ``hamiltonian'' $f$
is determined by
\be
{\cal L}_{X_f} g = \{g,f\}
\ee
where $\{ , \}$ are the Poisson brackets. 
Hence the derivation 
\be
{\cal L}_{X_f}(x_i) =  \{x_i,f\}
\ee
is an infinitesimal APD determined by $f$
acting on the coordinate function $x_i$.
After quantization, 
this is replaced by the commutator 
\be
\d_f(x_i) =  [x_i,f]
\ee
which can therefore be interpreted as quantized infinitesimal APD on 
the fuzzy sphere. In this way, the APD's on the fuzzy sphere
can be identified with $SU(N)$. 

Now consider ``abelian'' gauge theory on $S^2_N$ as discussed in
Section \ref{sec:3}. It is well known (see e.g. \cite{lizzi-szabo}) that 
gauge transformations on noncommutative spaces are closely related 
to certain diffeomorphism groups.
Indeed according to the above discussion, the gauge transformations 
\be
B_i \to U^{-1} B_i U
\ee
with $U = \exp(i f(x)) \in SU(N)$
can be interpreted as APD acting on the $B_i$, viewed as 3 scalar
fields. This statement has no classical analog, and 
has nothing to do with the action of APD's on classical 
gauge fields $A_i$. The induced transformation 
\be
F_{ij} \to U^{-1} F_{ij} U
\label{F-trafo-U1}
\ee
of the field strength becomes trivial in the commutative limit,
since then $F_{ij}$ commutes with all functions. For finite $N$ however,
\eq{F-trafo-U1} is nontrivial even in the ``abelian'' case at hand, and
can be interpreted as action of the APD determined by 
$U = e^{i f(x)}$ on each component of $F_{ij}$ (which is again
different from the classical action of APD's on a tensor!). 
Hence gauge transformations are necessarily ``non-local'', and
cannot be disentangled from general
coordinate transformations on $S^2_N$.
This phenomenon is quite common
on noncommutative spaces: e.g. on the 
canonical quantum plane $\R^d_\theta$, 
the translations
are inner and hence part of the gauge transformations of $F_{ij}$. 
One could interpret
this from a fiber-bundle point of view by saying that base space and fiber 
become unified in some sense. 

Apart from this $SU(N)$ group of gauge transformations (or APD's),
the constraint $C_0 = \frac 12$ is also preserved by the $SU(2)$ group 
$\exp(i \a_i \s^i)$ which acts on the indices of the $B_i$ only. 
The overall symmetry group is therefore  $SU(N) \times SU(2)$.
Note that the ``physical rotations'' 
$\exp(i \a_i (\la^i + \frac 12\sigma^i))$
are combinations 
of these $SU(2)$ rotations and gauge transformations.

Let us extract some information about the 
correlation functions using these symmetries,
without trying to calculate them explicitly. We only consider 
the abelian case for simplicity. 
The correlation functions are defined by
\beq
\langle C_1 ... C_m \rangle 
 = \frac 1Z\;\int dC  \; C_1 ... C_m\; 
   \d(C_0 -  \frac 12) \exp(-\Tr V(C)) 
\label{B-constr}
\eeq
where the indices $1, ..., m$ indicate $2N\times 2N$ matrix labels,
or in terms of the components
\beq
\langle (B_i)_1 ... (B_j)_m \rangle = \frac 1Z\;
  \int dB \; (B_i)_1 ... (B_j)_m\;  \exp(-\Tr V(B)).
\label{C-constr}
\eeq
They are highly constrained by the $SU(N) \times SU(2)$ symmetry. 
For example,
\be
\langle B_i \rangle  = 0
\label{exp-B}
\ee
using $SU(2)$ invariance.
This might seem strange, because we were using an expansion
$B_i = \la_i + A_i$ with $A_i$ being a ``small'' fluctuation.
However, this is not a contradiction: we do {\em not} assume any
spontaneous symmetry breaking (or gauge fixing), 
and the solution
$B_i = \lambda_i$ is only one possible gauge choice.
To get nontrivial results, we should of course consider 
quantities which are gauge invariant or contain gauge invariant 
information.
Consider for example
\beq
\langle F_i(x) F_j(y) \rangle = c\; \d_{ij} \; g_{ab}\; \nu^a \tens \nu^b
\eeq
using the $SU(N) \times SU(2)$ symmetry, where $\nu^a$ denotes the 
$SU(N)$ Gell-Mann matrices. Here $x$ and $y$ stand for the first respectively
second tensor slot, interpreted as functions on $S^2_N$. 
Since $g_{ab}\; \nu^a \tens \nu^b$
is the reproducing kernel, it should be interpreted as delta-function $\d(x,y)$
on the sphere, and we can write 
\beq
\langle F_i(x) F_j(y)\rangle = c \; \d_{ij} \; \delta(x,y).
\eeq
This makes sense: there are no propagating modes, therefore
there is no correlation between fields at different
points. The normalization can be calculated e.g. for coinciding points,
$\int \langle F_i(x) F^i(x)\rangle$
which is certainly nonzero but finite. 
Similarly, it follows that
\beq
\langle F_i(x) F_j(y) F_k(z) \rangle = \vare_{ijk} f_{abc} 
\nu^a \tens \nu^b \tens \nu^c
\eeq
where $f_{abc}$ is the structure constant of $SU(N)$,
since the r.h.s. is the only invariant tensor ($d_{abc}$ cannot
occur since the lhs is totally symmetric). Hence it is proportional
to the function 
\beq
f(x,y,z):= f_{abc} \nu^a \tens \nu^b \tens \nu^c
\eeq
on $S^2_N$, which in the classical limit is the unique function 
$f(x,y,z)$ on the sphere which is invariant under APD's and vanishes for
coinciding points. 

In general, one would expect that all correlation functions 
of the field strength have a 
well-defined classical limit in this model, 
just like the partition function. In
principle it should be possible to calculate them
explicitly in terms of
integrals over eigenvalues, and we expect no problem related 
to UV/IR mixing or renormalization. It would also be interesting to
know whether it is possible to relate them to integrable models.
These issues are left for further investigations.

\section{The $q$-deformed fuzzy sphere revisited}

Remarkably, we can repeat the same construction with a slightly
different potential, and obtain a gauge theory on the 
$q$-deformed fuzzy sphere \cite{qFSI}. 
Consider
\be \label{C0q}
C^{(q)} = \frac{1}{2} +  \l^{(q)}_i \s_{(q)}^i
\ee
where $\l^{(q)}_i$, $i=1,2,3$  are $N \times N$ hermitian matrices 
and $\s_{(q)}^i$ are the $q$-deformed sigma matrices,
both of which are defined to be the Clebsches of the corresponding
irreps of $U_q(su(2))$. 
To simplify the notation, we shall sometimes omit the label $(q)$, which is 
understood throughout this section. We also need the $q$-deformed
invariant tensors $\vare^{ij}_k = (\vare_{(q)})^{ij}_k$ and 
$g_{ij} = g_{ij}^{(q)}$ which can be found e.g. in \cite{qFSI}.
The generators satisfy 
\be \label{FSq-soln}
\vare^{ij}_k \l_i \l_j = i \l_k,
 \quad \l_i \l_j g^{ij} = \frac{[N-1]_q [N+1]_q}{[2]^2_{q^N}}
\ee
where $[n]_q = \frac{q^n-q^{-n}}{q-q^{-1}}$, and 
\be \label{q-sigma}
\vare_{ij}^k \s^i \s^j = i [2]_{q^2} \s^k,
 \quad \s^i \s^j g_{ij} =  [3]_q
\ee
which implies
\be
\s^i \s^j = i \vare^{ij}_k  \s^k + g^{ij}.
\ee 
It follows that
\be
(\l^{(q)}_i \s_{(q)}^i)^2 = - (\l^{(q)}_i \s_{(q)}^i) 
  + \frac{[N-1]_q [N+1]_q}{[2]^2_{q^N}}.
\ee
This means that $C^{(q)}$ has eigenvalues 
$\pm \sqrt{\frac{[N-1]_q [N+1]_q}{[2]^2_{q^N}} + \frac 14}$
with multiplicities $N\pm 1$. It is therefore a minimum of the matrix
potential
\be \label{Vq}
V_q(C) =\frac{1}{g^2 N} \left(C^2 
  -(\frac{[N-1]_q [N+1]_q}{[2]^2_{q^N}} + \frac 14) \right)^2.
\ee
Expanding now a general matrix $C$ as
\be
C =  \frac{1}{2} +  B_i \s_{(q)}^i, \qquad B_i = \l_i + \cA_i
\ee
imposing again the constraint $C_0 = \frac 12$,
one obtains 
\be
C^2 -(\frac{[N-1]_q [N+1]_q}{[2]^2_{q^N}} + \frac 14) 
 = (B_i B_j g^{ij} - \frac{[N-1]_q [N+1]_q}{[2]^2_{q^N}}) + \s^i F_i
\ee
where 
\be 
F_i = B_i + i\vare_i^{kl} B_k B_l, \quad
F_{kl}:= i [\l_k,\cA_l] - i [\l_l,\cA_k] + i [\cA_k,\cA_l] + \vare_{klm} \cA^m
\label{F-def-q}
\ee
is indeed the appropriate $q$-field strength as used in \cite{qFSI}.
Hence we naturally recover gauge models on the $q$-deformed fuzzy
sphere.
However it is not clear how to define the trace: If we 
take the classical trace 
\beq
S = \Tr\; V_q(C),
\eeq 
a strange term of the form $\Tr((B_iB^i) F_3)$ appears, because 
$\s^3_{(q)}= \left(\begin{array}{cc} q & 0 \\ 0 & -q^{-1}\end{array}\right)$ 
has a non-vanishing trace:
\be
tr (\s^i_{(q)} \s^j_{(q)}) = 2\; g^{ij}_{(q)} + i(q-q^{-1}) \vare_3^{ij}.
\ee
Hence it seems more natural to take the quantum trace over the 
$\s^i$ space, which has the property that 
$tr_q(\s^i_{(q)}) =0$, so that
$tr_q(\s^i_{(q)} \s^j_{(q)}) = [2]_q\; g^{ij}_{(q)}$.
Then taking the classical trace over the $N\times N$ matrices would 
lead to the action 
\beq
S = \frac{2}{g^2 N} \Tr\Big((B_i B^i  - \frac{N^2-1}4)^2 
+ (B_i + i\vare_{ijk}B^j B^k)(B^i + i\vare^{irs}B_r B_s)\Big).
\eeq
where $q$-deformed tensors are understood.
This is again invariant under $SU(N)$ (hence solvable), 
but the ``physical'' rotations defined similar as in 
Section \ref{sec:symmcorr} are violated.
On the other hand, 
taking the quantum trace over the full $2N\times 2N$ matrices
would break the gauge invariance but make the model
formally $U_q(su(2))$ invariant as in \cite{qFSI}. 
This may shed some light on the issues raised in the application of
$q$-deformed gauge theories on $D$-branes, see \cite{jacekme}.

\section{Discussion and outlook}

We presented a new formulation of pure gauge theory on the fuzzy 
sphere, which allows to carry out the path integral explicitly. 
The partition function for $U(n)$ Yang-Mills theory on 
$S^2_N$ is calculated in the large $N$ limit, and 
the known result \cite{migdal,rusakov} 
for the classical sphere is recovered for large $N$. 

There are several messages that should be stressed. First and foremost, 
gauge theories on noncommutative spaces are accessible to new 
methods and tools which could not be applied on commutative space.
Of course, Yang-Mills theory on $S^2$ is a rather simple field
theory with no propagating degrees of freedom; however it does have
nontrivial monopole sectors, and its quantization is not entirely trivial.
If the methods presented here 
can be generalized, noncommutative gauge theory
may become a useful alternative to  lattice gauge theory, even from an 
analytical point of view. Of course it 
should also be interesting from a numerical point of view. 

Another important message is that the
``classical limit'' of pure gauge theory on the fuzzy sphere is smooth,
at least for the partition function. This 
is not obvious in view of the UV/IR mixing effects
in noncommutative field theories. It would be very interesting to know
if this generalizes to higher dimensions. 
An interesting such space is  fuzzy $\C P^2$ \cite{cp2}, 
which is the subject of current investigations.

Furthermore, this new formulation of gauge theory 
on the fuzzy sphere is arguably {\em simpler\/} than on 
the classical sphere. It is defined by a potential 
for a hermitian matrix plus a constraint.
This leads not only to the correct kinetic terms, 
but also all the monopole sectors arise ``automatically'' 
in a very simple way, 
with the correct classical limit. In particular,  
there is no need to introduce nontrivial fiber bundles,
connections and other mathematical structures in our approach.
What is missing so far is the inclusion of fermions in this formalism; 
this will be discussed elsewhere.

There are several other aspects which require further work. First,
one should be able to calculate the correlation functions 
or other suitable observables for pure gauge theory.
Perhaps there are some connections with integrable models.
One may also try to extend this approach to other gauge groups.
Of course it would be very desirable to simplify the calculation in 
Appendix B, 
and to systematically calculate the corrections for finite $N$. 
Furthermore, there exist other interesting 
gauge models with similar action but 
without the term $(B_i B^i  - \frac{N^2-1}4)^2$, as discussed in \cite{ars1}.
They do not fit very well into the formalism presented here. 
However then the ``radial'' field $\varphi$ \eq{t1} becomes dynamical,
and these models describe branes on $SU(2)$ near the origin rather than
a Yang-Mills gauge theory on a sphere.
 All these
questions certainly deserve further study.

\section*{Acknowledgements} 

First and foremost I want to thank Chong-Sun Chu 
for many discussions, collaboration on several aspects of this paper 
and an invitation to Durham; in particular some 
of the results in Section 3 were obtained together with him.
I also wish to thank H. Grosse for his interest, encouragement and 
extensive discussions as well as initiations to the ESI in Vienna. 
Furthermore I am indebted to J. Wess, 
P. Aschieri, B. Jurco and P. Schupp for drawing my attention to 
noncommutative gauge theories, 
and I enjoyed useful discussions with E. Langmann, J. Madore, 
S. Rajeev, and P. Presnajder.

\section*{Appendix A. Some useful formulae}

The irreducible  $N$-dimensional representation of the 
$su(2)$ algebra $\l_i$ \eq{FS-l} is given by
\bea \label{reps}
&&(\l_3)_{kl} = \d_{kl} \; \frac{N+1-2k}{2},\\
&&(\l_+)_{kl} = \d_{k+1,l}\sqrt{(N-k)k},
\eea
where $k,l = 1, ..., N$ and  $\l_\pm = \l_1 \pm i\l_2$. 

Furthermore, recall that 
\be
\sigma^i \sigma^j = \d^{ij} + i \vare^{ijk} \sigma^k.
\ee
Together with \eq{FS-l}, this implies 
the following crucial property of the matrix $\la_i \sigma^i$:
\beq
(\la_i \sigma^i){}^{\;2} 
 =  \frac{N^2-1}{4} - (\la_i \sigma^i),
\label{lasi2}
\eeq
which means that the eigenvalues (in $Mat(2N)$) of the matrix
$\la_i \sigma^i$  are $\frac{-1 \pm N}{2}$. To get the multiplicities,
we note that $\la_i \sigma^i$ is an intertwiner of
$(2)\otimes(N) = (N-1) \oplus (N+1)$
(i.e. it is invariant under $SU(2)$), hence the
multiplicities are $N+1$ resp. $N-1$.

\section*{Appendix B: Evaluation of the partition function.}

Consider the expression
\be
\frac{\det(e^{i \L_i J_j})}{\D(J_i)}
\label{det-fraction}
\ee
in \eq{U-int}.
At first sight it may appear ill-defined, because the denominator 
is singular due to the form (\ref{J}) of $J$. However, the fraction is 
analytic in $J$, because
all poles are canceled by zeros in the determinant
(it must be so, because the lhs of \eq{U-int} is clearly analytic).
To see this explicitly, assume the following ``regularization''
\be  
\cJ = \left(\begin{array}{cc} K +\e\one & 0 \\ 0 &  K  \end{array}\right)
\ee
for some infinitesimal constant  $\e$, and
consider $\D(\cJ_j)$ in more detail. By treating the contributions
from the 2 blocks as above separately, it is easy to see that 
\be
\D(\cJ_j) = \D^4(k_i) \e^M.
\label{VDMJ}
\ee 
where $k_i$ are the eigenvalues of $K$. It is useful to order the $\L$ as
$\vec \L = (\L^{(+)}_1,\L^{(+)}_2, ..., \L^{(+)}_{n_+},\\
\L^{(-)}_1,\L^{(-)}_2, ..., \L^{(-)}_{n_-})$
where $\L^{(\pm)}_i \approx \pm\frac N2$.
We then have to evaluate the determinant
\beq
\det(e^{i \L_i \cJ_j}) =
\det\(\begin{array}{cc} e^{i \L^+_{i} (k_j + \e)}, &  e^{i \L^+_{i} k_j}  \\
     e^{i \L^-_{i} (k_j+ \e) } , &   e^{i \L^-_{i} k_j}\\
     \end{array}\),
\label{detLJ}
\eeq
which clearly contains a factor $\e^M$ due to the degeneracy in the
$\{k_i\}$. To proceed, we expand it by choosing 
$M+n$ resp. $M-n$ columns in the upper resp. lower block of \eq{detLJ}
as follows
\beq
\det(e^{i \L_i \cJ_j}) = \sum_{\{\cJ^+\},\{\cJ^-\}} (-1)^{\sigma(\{\cJ^+\},\{\cJ^-\})}
  \det\big( e^{i \L^+_{i} \cJ^+_j}\big)
  \det\big( e^{i \L^-_{i} \cJ^-_j}\big)
\eeq
where $\{\cJ^+\},\{\cJ^-\}  \subset \{k_i+\e,k_i\}$ are complementary subsets 
with $|\cJ^+| = M+n, |\cJ^-| = M-n$, and 
the sign is given by this choice of subsets.
In the terms $ \det\big( e^{i \L^+_{i} \cJ^+_j}\big)$, the $\cJ^+$
are assumed to be ordered as in $(k_1+\e, ...,k_M+\e,k_1, ..., k_M)$, 
so that only the 
choice of the sub{\em sets} $\{\cJ^{\pm}\}$ matters for the sign.
In terms of the fluctuations 
\bea
\mu_j^- &=& (\L_j^- +\frac N2), \quad j = 1,2,..., M-n, \nn\\
\mu_j^+ &=& (\L_j^+ -\frac N2), \quad j = 1,2,..., M+n
\eea
this becomes 
\beq
\det(e^{i \L_i \cJ_j}) = \sum_{\{\cJ^+\},\{\cJ^-\}} (-1)^{\sigma(\{\cJ^+\},\{\cJ^-\})}
  \det\big( e^{i \mu^+_{i} \cJ^+_j}\big)
  \det\big( e^{i \mu^-_{i} \cJ^-_j}\big)\; 
   e^{i \frac N2 (\sum \cJ^+ - \sum \cJ^-)}.
\label{detLJ-2}
\eeq
Now the rapidly oscillating terms have been isolated in the last exponential,
and it turns out that the correct expansion is in the number of
these rapidly oscillating variables $k_i$ in $e^{i \frac N2 \cJ}$. 
This depends on the choice of $\{\cJ^+\}$ (which of course fixes $\{\cJ^-\}$):
Let $\{\k^+\}$ be the set of $k_i$'s which occur twice in $\{\cJ^+\}$
(for $\e=0$). Because $|\cJ^+| = M+n$, there are at least $n$ such $\k^+$'s.
Assume that $|\{\k^+\}| = n+d$: then there must be in addition $d$ elements 
$\k^-_j$ among the $\{k_i\}$ which occur twice in $\{\cJ^-\}$.
The last term in \eq{detLJ-2} is then $e^{i N (\sum \k^+ - \sum \k^-)}$,
which means that there are $n+2d$ rapidly oscillating variables among 
the $k_i$. We can therefore expect that $d=0$ will give the dominating
contribution, and concentrate on this case first. Then 
$\{\cJ^+\} = \{\cJ^-\} \cup 2\{\k^+\}$ for $\e =0$, and 
\beq
\det(e^{i \L_i \cJ_j})_{d=0} =
   \sum_{\{\cJ^+\}, \{\cJ^-\};\; d=0} (-1)^{\sigma(\{\cJ^+\},\{\cJ^-\})}
  \det\big( e^{i \mu^+_{i} \cJ^+_j}\big)
  \det\big( e^{i \mu^-_{i} \cJ^-_j}\big)\;  e^{i N (\sum \k^+)}.
\eeq
We will omit the superscript of $\k$ from now on. 
Consider $\{\cJ^+\}$ in more detail, for fixed $\{\k\}$. It has the form
$\{\k\} \cup \{\k+\e\} \cup \{\cJ^- \pm \e\}$, and each different sign
choice in $\{\cJ^- \pm \e\}$ gives a different contribution 
to the sum which must be added. 
Different choices are related by an exchange of the elements 
$k_i, k_i+\e \not \in \{\k\}$ in 
$\cJ^\pm$. This amounts to an exchange of the corresponding columns
in \eq{detLJ}, hence the two contributions come with a relative $-$
sign. Among these, there is one ``ordered'' choice
$(\cJ^+)_0 = (k_1+\e, ..., k_M+\e,  \k_1, ..., \k_n)$, $(\cJ^-)_0 =(k_i')$,
which depends only on $\{\k\}$ and serves to fix the sign 
$(-1)^{\sigma(\k)}:= (-1)^{\sigma((\cJ^+)_0,(\cJ^-)_0)}$.
The following notation is useful:
\beq
T^\e_i f(k_1, ..., k_M) = f(k_1, .., k_i+\e, ..., k_M)
\eeq
so that e.g. 
\beq
(\cJ^+)_0 = \prod_{i=1}^M T^\e_i (k_1, ..., k_M,  \k_1, ..., \k_n)
 =  \prod_i T^\e_i J^+
\eeq
where $J^+ = (k_1, ..., k_M,  \k_1, ..., \k_n)$, 
$J^- = (k'_1, ..., k'_{M-n})$ from now on.
Summing over all these permutations for each set $\{\k\}$ leads to 
\beq
\det(e^{i \L_i J_j})_{d=0} =
   \sum_{\{\k\}} (-1)^{\sigma(\k)} e^{i N (\sum \k)}
   \prod_{i \not\in \{\k\}} \Big((T^\e_i)_+ - (T^\e_i)_-\Big) 
   \det\big( e^{i \mu^+_{i} J^+_j}\big)
  \det\big( e^{i \mu^-_{i} J^-_j}\big).
\label{detLJ-3}
\eeq
Here $(T^\e_i)_\pm$ indicates that the $T^\e_i$ operator acts only on 
$\det\big( e^{i \mu^\pm_{i} J^\pm_j}\big)$.
The sign $(-1)^{\sigma(\k)}$ now depends 
only on the choice of $\{\k\} \subset \{k_i\}$.
Using 
\beq
(T^\e_i)_+ - (T^\e_i)_-  = 
  ((T^\e_i)_+ -1) - ((T^\e_i)_- -1)
  \to \e \Big((\partial_{k'_i})_+ - (\partial_{k'_i})_-\Big)
\eeq
for $\e \to 0$, we get 
\beq
\det(e^{i \L_i J_j})_{d=0} =
  \e^{M-n} \sum_{\{\k\}} (-1)^{\sigma(\k)}\; e^{i N (\sum \k)}
   \prod_{\{k'\}}  \Big((\partial_{k'_i})_+ - (\partial_{k'_i})_- \Big) 
   \det\big( e^{i \mu^+_{i} J^+_j}\big)
  \det\big( e^{i \mu^-_{i} J^-_j}\big).
\label{detLJ-4}
\eeq
We have hence recovered most of the $\e$ factors; the missing $\e^n$ 
is contained in $\det\big( e^{i \mu^+_{i} J^+_j}\big)$. 

To proceed, we can now integrate over $\mu^\pm$ in \eq{Z-J2},
noting that 
\beq
\Delta(\L) = N^x\;\D(\mu^+) \D(\mu^-) 
  \exp(n(\Sigma \mu^{-} - \Sigma\mu^{+}))
\label{Deltamu-approx}
\eeq
up to corrections of order $1/N$. In fact we can even neglect the last
exponential compared to the leading terms 
$\exp(-\frac N{g^2} \sum (\mu^\pm)^2)$
in the action, expanded as in \eq{V-expand}. 
Furthermore,
\beq
\int d\mu^+ \D(\mu^+) \det(e^{i \mu^+_{i} J^+_j})\;
\exp(-\frac N{g^2} \sum (\mu^+)^2) = c\; \D(J^+)\; 
\exp(-\frac{g^2}{4N} \sum (J^+)^2)
\eeq
for some constant $c$ by antisymmetry in $J^+$, and
similarly for $\mu^-$.
Noting also that 
\beq
\D(J^+) \D(J^-)=
 \Big((-1)^{\sigma(\k)}\e^n \D^4(\k)\D(k')\prod(k'-\k)^2\Big)_+\; 
 \Big(\D(k')\Big)_-  
\eeq
we get using \eq{VDMJ}
\beqa
Z_{d=0} &=& 
 \int dk_i\; \frac{e^{-i \sum k_i}}{\D(k)^2}\; 
 \sum_{\k} \; \D^4(\k) \exp\( i N (\sum \k) -\frac{g^2}{2N}(\sum \k^2)\)\nn\\
 && \!\!\! \!\!\!\!\!\!  \!\!\!\prod_{\{k'\}}  
      \Big((\partial_{k'_i})_+ - (\partial_{k'_i})_- \Big)
      \Big(\prod(k' -\k)^2\D(k') \exp(-\frac{g^2}{4N}\sum k'^2) \Big)_+
   \Big(\D(k') \exp(-\frac{g^2}{4N}\sum k'^2) \Big)_- \nn
\eeqa
always ignoring numerical constants.
Notice that all signs have disappeared. Since
\beq
\big(\partial_{k_1} - \partial_{l_1}\big)(f(k_1) f(l_1))|_{k_1=l_1} =0,
\eeq
only the term 
$\partial_{k'_1}\prod(k' -\k)^2 = (\sum_i \frac 2{k'_1 - \k_i})\prod(k' -\k)^2$
survives the derivatives w.r.t. $k'_1$, and the term 
$(\sum_i \frac 2{k'_1 - \k_i})$ can be moved outside of the remaining
derivatives\footnote{If we were more
careful to include the factor $\exp(n(\Sigma \mu^{-} - \Sigma \mu^{+}))$, 
there would be additional contributions which are suppressed by $\frac 1N$}. 
Repeating this, we find 
\beqa
Z_{d=0} &=& 
 \int dk_i\; \frac{e^{-i \sum k_i}}{\D(k)^2}\; 
 \sum_{\k} \; \D^4(\k) \D^2(k') e^{i N (\sum \k)}
 \exp(-\frac{g^2}{2N}\sum k^2 )  
   \(\prod_{\{k'\}} \partial_{k'_i} \prod(k' -\k)^2\)\nn\\
 &=& 
  \int dk_i\; e^{-i \sum k_i}\; 
 \sum_{\k} \; \D^2(\k) e^{i N (\sum \k)}
 \exp(-\frac{g^2}{2N}\sum k^2) \(\sum \prod\frac 2{k' -\k}\)
\eeqa
since $\D^2(k) = \D^2(\k)\D^2(k')\prod(k'-\k)^2$. Here
\beq
\sum \prod\frac 1{k' -\k} = 
  \sum (\prod^{N_1}\frac 1{k' -\k_1}) ... (\prod^{N_n}\frac 1{k' -\k_n})
\eeq
is the sum over all possible splittings of $\{k'\}$ into $n$ partitions
$\{k'\}_{N_1}, ..., \{k'\}_{N_n}$ of size $N_1 + ... + N_n = M-n$,
and 
$(\prod^{N_i}\frac 1{k' -\k_i})\equiv(\prod_{\{k'\}_{N_i}}\frac 1{k'  -\k_i})$.
That is, each $k'_i$ is ``linked'' to one $\k_j$ by a factor of the form
$\frac 1{k'_i -\k_j}$, and there are $N_j$ such $k'$ linked to $\k_j$.
For fixed $(N_1, ..., N_n)$ each of these terms gives the same
contribution, therefore we simply get a multiplicity factor 
$\frac{(M-n)!}{N_1! ... N_n!}$. Furthermore
each different choice of  $\{\k\} \subset \{k\}$ gives the same integral
since the form is identical. Therefore
\beqa\label{Z-5}
Z_{d=0} &=& 
\int dk_i\; e^{-i \sum k_i + i N (\sum \k) }\;\D^2(\k) \nn\\
&&  \!\!\! \!\!\!\!\!\! \!\!\sum_{N_1 + ... + N_n = M-n}\; 
\frac{(M-n)!}{N_1! ... N_n!} 
  \prod^{N_1}(\frac 1{k' -\k_1}) ... \prod^{N_n}(\frac 1{k' -\k_n})
 \exp(-\frac{g^2}{2N}\sum k^2).
\eeqa
We have now taken just one fixed subset $\{\k\} \subset \{k\}$, since all
choices give the same result. 

The integral over the $k'_i$ has now the same structure for all $i$ and
can be carried out. Notice that there
are poles in \eq{Z-5}, which seems a bit surprising because the
original expression $\frac{\det(e^{i \L_i J_j})}{\D(\L_i) \D(J_j)}$ 
is perfectly regular. However, recall that this corresponds
to the sum over all possible terms in \eq{detLJ-2}, in particular over
all choices of $\{\k \} \subset \{ k\}$. In \eq{Z-5} these different
contributions are included in the 
integral over all values $\int_{-\infty}^{\infty} dk_i$, contributing 
with the same $|k_i-\k_j|$ but opposite sign.
Therefore the cancellations in the sum over $\{\k\}$
are now reflected in the cancellations in the 
contributions to the integral from both sides of the poles 
{\em with the same $|k_i-\k_j|$}. We should therefore use the principal
value of the integral $\Pint d k_i$ in \eq{Z-5}, which 
is perfectly regular and well-defined because there are only simple poles. 
To put it differently, we can restrict the range of integration to 
the space of $k_i$ with $|k_i - k_j| < \e'$, say; this must give the correct
result for $\e' \to 0$. But this is just the definition of the
principal value of the integral.
In fact, notice that if there were no poles in
the above formula, each integral $\int d k'_i$ would produce an
exponential factor of order $e^{-N/g^2}$, and $Z$ would vanish
for large $N$. The contributions from the poles will give the correct,
finite result. 

Using the identity
\beq
\Pint d u\; \frac 1{u} f(u) = \frac 12 \(\Pint d u\; \frac{f(u)-f(-u)}{u}\)
\eeq
we get 
\beqa
\Pint  d k'\; \frac 1{k' -\k}&& \!\!\!\!\!\!\!\!\!\!\!\! 
          \exp(-i k' -\frac{g^2}{2N}(k')^2) =\nn\\
 &=&  e^{-i \k -\frac{g^2}{2N}\k^2 }\;
   \Pint d u\; \Big( -i\; e^{-u\frac{g^2}{N}\k}\;\frac{\sin(u)}{u} 
      -  e^{iu}\; \frac{\sinh(\frac{g^2}{N}\k u)}{u}\Big)
    \; \exp(-\frac{g^2}{2N}u^2) \nn
\eeqa
Now $\frac 1u \sin(u) \approx \pi \d(u)$, and 
$\frac 1u \sinh(\frac{g^2}{N}\k u) \approx 0$ under the integral, hence
\beq
\Pint  d k'\; \frac 1{k' -\k}\exp(-i k' -\frac{g^2}{2N}(k')^2) 
 \to -i\pi\; e^{-i \k -\frac{g^2}{2N}\k^2 } \qquad \mbox{for large } N.
\label{principal-result}
\eeq
We therefore obtain
\beq
Z_{d=0} = 
 \sum_{N_1 + ... + N_n = M-n}\; 
      \frac{(M-n)!}{N_1! ... N_n!}
   \int d \k_i\; e^{i \k_i (N - (N_i+1))}\;\D^2(\k) 
    \exp(-\frac {g^2}{2}\sum \k_i^2 (\frac{N_i+1}{N})).
\label{Z-6}
\eeq
In terms of the integers $m_i = N-(N_i+1)$  
which satisfy 
\beq
\sum m_i =nN -M =: m,
\eeq
the combinatorial factor is 
\beq
\frac{(M-n)!}{N_1! ... N_n!} \approx \frac{(M-n)!}{((\frac Mn -1)!)^n}
\label{combin-approx}
\eeq
as long as the $m_i$ are small (which will be justified below),
up to corrections of order $\frac 1N$. Hence we can drop this factor, 
and obtain
\beq
Z_{d=0} = 
\sum_{m_1 + ... + m_n = m}\; 
     \int d \k_i\;\D^2(\k) \;  e^{i \k_i m_i}\;
    \exp(-\frac {g^2}{2}\sum \k_i^2 (1 - \frac {m_i}N)).
\label{Z-7}
\eeq
This is now a perfectly nice integral. The $ \frac {m_i}N$
in the exponential can be neglected, and \eq{Z-n-nice} follows.

Several remarks are in order. 

\bit

\item If one would similarly treat the case of more oscillating factors
$d >0$ in \eq{detLJ-2}, we would get similar formulas with $\k$ replaced by 
$n+2d$ variables $\k^+, \k^-$, which come with oscillating terms 
$e^{iN \k_i^\pm}$. The remaining analysis would be similar, 
with only $N-n-2d$ variables $k'$ which contribute an integral
as in \eq{principal-result}. This leads to an expression 
similar as in \eq{Z-6}, however there are now not enough phase factors
$e^{- i\k_i^\pm}$ to cancel the rapid oscillations in 
$e^{iN \sum \k_i^\pm}$. Therefore this  integral 
will have additional rapidly oscillating terms, which 
lead to an exponential suppression. Therefore the contributions
$Z_{d > 0}$ can be neglected for large $N$.

\item The approximation replacing the
 combinatorial factor  in \eq{combin-approx} 
 by a constant is justified, since only small $|m_i|$ 
contribute to the final result \eq{Z-9}. 

\item one can understand the above calculations, in particular the 
dominant contribution from the poles in \eq{Z-5} intuitively as
follows: Each contribution from the poles with 
$(N_1, ..., N_n)$ corresponds to a ``clustering'' 
$(\{k'\}_{N_1}\approx\k_1), ..., (\{k'\}_{N_n}\approx\k_n)$,
which gives the dominant contribution to the integral over the $k_i$. 
These clusters correspond precisely to the saddle-points discussed in Section 
\ref{sec:nonabelian},
which are also clusters of $N_i+1$ eigenvalues of $C$.
Hence the leading contribution comes form the saddle-points. 

\item This calculation could be generalized to the potential
  \eq{Vprime}, ordering the eigenvalues as 
  $\vec \L = (\L^{(+)}_1,\L^{(+)}_2, ..., \L^{(+)}_{n_+}, \L^{(0)}_1,
  ..., \L^{(-)}_1,\L^{(-)}_2, ..., \L^{(-)}_{n_-})$
where $\L^{(\pm)}_i \approx \pm\frac N2$ and 
$\L^{(0)}_i \approx \frac12$.
This would allow to explicitly calculate the relative weights
of the different topological sectors in \eq{Z-full}, which have been put
by hand here.

\item with some effort, it should also be possible to compute 
the leading correction terms in $\frac 1N$. The relevant
approximations are those in \eq{combin-approx} and
\eq{principal-result}, which can certainly be improved. The other
approximations (taking only $d=0$, ignoring the exponential in 
in \eq{Deltamu-approx} an the higher terms in \eq{V-expand})
apparently give corrections which are exponentially suppressed.
An exact calculation is certainly desirable, but
would require more sophisticated tools.

\item needless to say, it would be nice to simplify this calculation.

\eit

\end{document}